\documentclass[12pt]{article}

\usepackage{cite,amsmath,amsfonts,amsthm,fullpage}
\usepackage{amssymb}
\usepackage{amsbsy}
\usepackage{epsfig}
\usepackage{verbatim}
\usepackage{amssymb}
\usepackage{amsbsy}
\usepackage{epsfig}
\usepackage{verbatim}
\renewcommand{\theequation}{\arabic{section}.\arabic{subsection}.\arabic{equation}}
\makeatletter \@addtoreset{figure}{section}
\def\thefigure{\thesection.\@arabic\c@figure}
\def\fps@figure{h, t}
\@addtoreset{table}{bsection}
\def\thetable{\thesection.\@arabic\c@table}
\def\fps@table{h, t}
\@addtoreset{equation}{section}
\def\theequation{\thesection.\arabic{equation}}
\newtheorem{corollary}{Corollary}[section]
\newtheorem{definition}{Definition}[section]

\newtheorem{proposition}{Proposition}[section]

\newtheorem{examps}{Examples}[section]

\newtheorem{lemma}{Lemma}[section]
\newtheorem{remark}{Remark}[section]
\newtheorem{remarks}[remark]{Remarks}

\def\bx{\begin{example}}
\def\ex{\end{example}}
\def\bxs{\begin{examps}. \rm\begin{enumerate}}
\def\exs{\end{enumerate}\end{examps}}
\def\bd{\begin{definition}}
\def\ed{\end{definition}}
\def\bp{\begin{proposition}\rm}
\def\ep{\end{proposition}}
\def\bc{\begin{corollary}}
\def\ec{\end{corollary}}
\def\bl{\begin{lemma}\em}
\def\el{\end{lemma}}
\def\be{\begin{equation}}
\def\ee{\end{equation}}
\def\br{\begin{remark}\rm\small}
\def\er{\end{remark}}
\def\brs{\begin{remarks}.\\ \rm\
\begin{enumerate}}
\def\ers{\end{enumerate}\end{remarks}}
\def\bea{\begin{eqnarray}}
\def\eea{\end{eqnarray}}

\def\ra{{\rightarrow}}

\def\det{\mathrm {det}}

\def\ds{\displaystyle}

\def\&{&{\hskip -20pt}}

\def\AA{{\mathcal A}}

\def\Zb{{\mathbf Z}}

\date{}

\begin{document}
\baselineskip 16pt
\begin{flushright}
CRM-3xxx(2006)
\end{flushright}
\medskip
\begin{center}
\begin{Large}\fontfamily{cmss}
\fontsize{17pt}{27pt} \selectfont \textbf{Fermionic approach to
the evaluation of integrals of rational symmetric functions}
\footnote{Work of (J.H.) supported in part by the Natural Sciences
and Engineering Research Council of Canada (NSERC) and the Fonds
FCAR du Qu\'ebec; that of (A.O.) by the Russian Academy of Science
program  ``Mathematical Methods in Nonlinear Dynamics" and  RFBR
grant No 05-01-00498.}
\end{Large}\\
\bigskip
\begin{large}  {J. Harnad}$^{\dagger \ddagger}$\footnote{harnad@crm.umontreal.ca}
 and {A. Yu. Orlov}$^{\star}$\footnote{orlovs@wave.sio.rssi.ru}
\end{large}
\\
\bigskip
\begin{small}
$^{\dagger}$ {\em Centre de recherches math\'ematiques,
Universit\'e de Montr\'eal\\ C.~P.~6128, succ. centre ville,
Montr\'eal,
Qu\'ebec, Canada H3C 3J7} \\
\smallskip
$^{\ddagger}$ {\em Department of Mathematics and Statistics,
Concordia University\\ 7141 Sherbrooke W., Montr\'eal, Qu\'ebec,
Canada H4B 1R6} \\
\smallskip
$^{\star}$ {\em Nonlinear Wave Processes Laboratory, \\
Oceanology Institute, 36 Nakhimovskii Prospect\\
Moscow 117851, Russia } \\
\end{small}
\end{center}
\bigskip
\bigskip

\begin{center}{\bf Abstract}
\end{center}
\smallskip

\begin{small}
We use the fermionic construction of two-matrix model partition
functions to evaluate integrals over rational symmetric functions.
This approach is  complementary to the one used in the paper
``Integrals of Rational Symmetric Functions,  Two-Matrix Models
and Biorthogonal Polynomials'' \cite{paper2}, where these
integrals were evaluated by a direct method.  Using Wick's
theorem, we obtain the same determinantal expressions in terms of
biorthogonal polynomials as in  \cite{paper2}.
\end{small}
\bigskip

\section{Introduction}

In this work, we consider the following integral
 \bea
 \label{mainintegral'}
 {\bf I}^{(2)}_N(\xi, \zeta, \eta, \mu) &:=& {1\over {\bf Z}^{(2)}_N } {\int\int}
d\mu(x_1,y_1)
 \dots {\int\int} d\mu(x_N,y_N)
 \Delta_N(x)\Delta_N(y) \cr
 && \quad  \times \prod_{a=1}^N {\prod_{\alpha=1}^{L_1}(\xi_\alpha - x_a)
 \prod_{\beta=1}^{L_2}(\zeta_\alpha - y_a) \over \prod_{j=1}^{M_1}
 (\eta_j - x_a)  \prod_{k=1}^{M_2}(\mu_k - y_a)}, \cr
 &&
 \eea
  where
  \be \Delta_N(x) := \prod_{i > j}^N (x_i-x_j), \quad
\Delta_N(y) := \prod_{i > j}^N (y_i-y_j)
 \ee
 are Vandermonde determinants, and
 \be
{\bf Z}^{(2)}_N :=\int\int d\mu(x_1,y_1)
 \dots \int\int d\mu(x_N,y_N)
 \Delta_N(x)\Delta_N(y),
 \ee
  is a normalization constant, which in the context of two-matrix models is
  interpreted as the partition function. Here $d\mu(x,y)$ is a measure (in general, complex),
  supported on a finite set of products of curves in the complex $x$ and $y$ planes.

It is known (see for instance \cite{CZ}) that this integral
can be presented in the form of a determinant of an $N\times N$
matrix; namely,
 \be
 {\bf I}^{(2)}_N(\xi, \zeta, \eta, \mu)
=
{N!\over {\bf Z}^{(2)}_N }\det\left( \int\int{\prod_{\alpha=1}^{L_1}(\xi_\alpha - x)
 \prod_{\beta=1}^{L_2}(\zeta_\beta - y) \over \prod_{j=1}^{M_1}
 (\eta_j - x)  \prod_{k=1}^{M_2}(\mu_k - y)}d\mu(x,y)
(x,y)x^jy^k \right)_{0\le j,k \le N-1}.
\label{detN-by-N}
 \ee
However for some purposes, it is more useful  to express it as the
determinant of a matrix whose size does not depend on $N$.

The problem of evaluation of integrals of such symmetric functions
of $N$ variables is of importance in the context  of matrix models
\cite{Mehta}, where $N$ is the number of eigenvalues. Expressing
integrals that determine correlation functions and expectation
values  as  determinants of matrices whose size  is independent of
$N$ is of importance e.g.,  in the study of $N\to\infty$ limits.
For one-matrix models such integrals were studied in
\cite{U},\cite{BH},\cite{FS},\cite{AV} and \cite{BDS}. In the case
of complex, normal or two-matrix models \cite{IZ} the problem was
considered in \cite{AV},  \cite{Be}, \cite{paper2}.

\br
In order to compare with  the results in \cite{AV}  and \cite{Be} on complex integrals
the variables $ \{x_a, y_a\}_{a=1, \dots N}$ must be replaced by complex conjugate pairs
$ \{z_a, \bar{z}_a\}_{a=1, \dots N}$, and the integration domains taken
as $N$ copies of the complex plane.
\er

The present paper is complementary to ref.  \cite{paper2}, where
such integrals were evaluated by a  ``direct'' method based on
partial fraction expansions and the Cauchy-Binet identity. Ref.
\cite{paper2} and the present work are intended to provide concise
presentations and new derivations of previously known as well as
new results, using two distinct methods.

  The approach used here is based on constructing a fermionic representation of integrals
   of such rational symmetric functions, introduced in \cite{paper1}, using two--component
     fermions. This representation is different from  the ones used previously in the
       context of  matrix models in \cite{ZKMMO}, \cite{HO1} and \cite{HO2},
which were based on one-component fermions.

\br
Although there is a close connection between matrix integrals, orthogonal polynomials and the
 spectral transform approach to the theory of integrable systems, in the present work we do not
   develop this latter aspect, which concerns the deformation theory
of the measures involved. Our starting point however is similar to
the previous paper \cite{paper1}, where the deformation theory is
addressed, and the fermionic approach to integrable systems
developed in \cite{DJKM} is utilized. The relation of these
results to the objects appearing in the spectral transform
approach is, however, briefly explained in the Appendices, where
the biorthogonal polynomials and their Hilbert transforms are
interpreted as Baker functions and adjoint Baker functions. \er

The language of free two-component fermions used here is borrowed
from \cite{DJKM}. The integrals that we are interested in are
expressed as vacuum state expectation values of operator products
formed from free fermion generators.  Besides successive
application of Wick's theorem, the main computational device used
consists of applying canonical (``dressing'') transformations to
pass from free fermions fields, whose Laurent coefficients are the
free Fermi creation and annihilation operators, to ``dressed'
ones, where the monomials are replaced by biorthogonal
polynomials. Wick's theorem then serves to express the same
operator product vacuum state expectation value (VEV) both as a
multiple rational integral and as  the determinant of a matrix
formed from elementary factors involving orthogonal polynomials
and their Hilbert transforms.

 Let $d\mu(x,y)$ be a measure (in general, complex), supported on a
finite set of products of curves in the complex $x$ and $y$
planes, for which the semi-infinite matrix of bimoments is finite:
 \be B_{jk} := {\int\int} d\mu(x,y)x^j y^k <\infty, \quad 0 , \quad
\forall j,k \in {\mathbb{N}}.
 \ee
 The integrals are understood to be evaluated on a specified linear combination of products of
 the support curves. Assuming that, for all $N\ge 1$, the $N \times N$
submatrix  $(B_{jk})_{0\le j,k, \le N-1}$ is nonsingular, the
Gram-Schmidt process may be used to construct an infinite sequence
of pairs of biorthogonal polynomials $\{P_j (x),
S_j(y)\}_{j=0\dots \infty}$, unique up to signs, satisfying
 \be
{\int\int} d\mu(x,y) P_j(x)S_k(y) = \delta_{jk},
 \ee
 and normalized to have leading coefficients that are  equal:
 \be P_j(x) =
{x^j\over\sqrt{h_j}} + O(x^{j-1}), \qquad S_j(x) = {y^j\over
\sqrt{h_j} }+ O(y^{j-1}).
 \ee
 We will also assume that  the Hilbert transforms of these biorthogonal polynomials,
 \be\label{HT}
\tilde{P}_n(\mu) := {\int\int} d\mu(x,y) {P_m(x) \over \mu -y},
\quad \tilde{S}_n(\eta) := {\int\int} d\mu(x,y) {S_m(x) \over \eta
-x},
 \ee
exist for all $n\in {\mathbb{N}}$.

The following expression for ${\bf Z}^{(2)}_N$ in terms of the leading term normalization
factors $h_n$ is then easily shown to hold (see, e.g.,  \cite{Mehta}).
 \be
 {\bf Z}^{(2)}_N =N!\det \left(B_{jk}\right)|_{j,k=1,\dots,N} =N! \prod_{n=0}^{N-1} h_n
 \label{partfn}
 \ee

Defining
 \be N_1:=N+L_1-M_1, \qquad N_2:=N+L_2-M_2,
  \ee
we consider different cases. In each case the answer is written in
form of the determinant of a matrix $G$ which is different for
different cases. Results are given

(1) $N_2\ge N_1\ge 0$: by formulae
(\ref{mainintegral-ref-1a})-(\ref{02a'-new}). $G$ is  a $(L_2+M_1)
\times (L_2+M_1)$ matrix

(2)  $N_1\le 0\le N_2$: by formulae
(\ref{answer-b-paper2})-(\ref{case2}). $G$ is a
$(L_2+M_1)\times(L_2+M_1)$ matrix

(3) $N_1\le N_2\le 0$:  by formulae
(\ref{answer-c-paper2*})-(\ref{case3}). $G$ is a
$(M_1+M_2-N)\times (M_1+M_2-N)$ matrix

The cases
 $N_1\ge N_2\ge 0$, $N_2\le 0\le N_1$ and
 $N_2\le N_1\le 0$
 are related to previous ones by  interchanging
 quantities $L_1\leftrightarrow L_2$, $M_1\leftrightarrow M_2$,
 $N_1\leftrightarrow N_2$, $\xi \leftrightarrow
\zeta$,  $\mu \leftrightarrow \eta$,  $P_n \leftrightarrow S_n$
and ${\tilde P}_n \leftrightarrow {\tilde S}_n$ in the final
formulae. (See respectively
(\ref{mainintegral-ref-1a})-(\ref{02a'-new}) vs.
(\ref{mainintegral-paper2})-(\ref{wtK21-paper2}),
(\ref{answer-b-paper2})-(\ref{case2}) vs.
(\ref{answer-b-paper2''})-(\ref{case2''}) and
(\ref{answer-c-paper2*})-(\ref{case3}) vs.
(\ref{answer-c-paper2''})-(\ref{case3''}).)

For instance, for the case  $N_1\ge N_2\ge 0$ the answer is
 \bea
 {\bf I}^{(2)}_N(\xi, \zeta, \eta, \mu)
 &=& (-1)^{{1\over 2}(M_1+M_2)(M_1+M_2-1)}(-1)^{L_1 M_2}
 \prod_{n=0}^{N-1}h_n^{-1}\prod_{n=N}^{{}_{N+L_2-M_2-1}}{\hskip -15 pt}\sqrt{h_n}
 \prod_{n=N}^{{}_{N+L_1- M_1-1}}{\hskip -15 pt}\sqrt{h_n} \cr
 &\& \quad \times
{ \prod_{\alpha=1}^{L_1}\prod_{j=1}^{M_1}(\xi_\alpha -
\eta_j)\prod_{\beta=1}^{L_2} \prod_{k=1}^{M_2}(\zeta_\beta- \mu_k)
 \over \Delta_{L_1}(\xi)  \Delta_{L_2}(\zeta)  \Delta_{M_1}(\eta)  \Delta_{M_2}(\mu)}
 \  \det G,  \cr
 &&
 \label{mainintegral-paper3-introd}
 \eea
where $G$ is a $(L_1+M_2)\times(L_1+M_2)$  matrix whose entries
 are expressed in terms of the values of $P_n,S_n,{\tilde P}_n$ and
 ${\tilde S}_n$ for $n=0,\dots,\max(N_1,N_2)$ evaluated at the points
 $\{\xi_\alpha, \zeta_\beta, \eta_j, \mu_k\}$.

 \br \label{rem-conditions}
  It is assumed that the support curves for the
 integrals involved do not pass through the points
 $\{\eta_j, \mu_k\}_{j=1,\dots M_1,  k=1,\dots M_2} $. The definition of the integrals
 may be extended throughout the complex plane of these parameters by analytic continuation, but
  might then take on  multiple values. However, if the measures $d\mu(x,y)$ vanish with sufficient
   rapidity as
 these points approach the support curves, the integrals considered
 in the present paper become single valued. If the replacements
$ \{x_a, y_a\}_{a=1, \dots N} \ra  \{z_a, \bar{z}\}_{a=1, \dots N}$,
are made, in order to compare with the results of \cite{AV}, \cite{Be},
it must be assumed that the corresponding measures $d\mu(z, \bar{z}$
in the complex plane vanish sufficiently rapidly at the points $\{z=\eta_j, \bar{z}=\mu_k\}$
for the integrals to converge.
   \er

In the following sections, we present the integral
(\ref{mainintegral'}) as a vacuum state expectation value of a certain
fermionic expression, using two-component fermions (see
(\ref{M-1-L1-M-2-L2-insertions}) below). To evaluate this
expectation value we use Wick's theorem, which expresses the
vacuum state expectation value of products of linear combinations of
fermions as a determinant of the matrix formed by evaluating
the pair-wise VEV's. Since the expression (\ref{M-1-L1-M-2-L2-insertions})
is not as yet of the form of a vacuum expectation value of a product of fermions, we use a set of
 tricks (namely, canonical transformations of fermions and re-writing of
 charged vacuum vectors in a suitable form)
 which serve also to reduce the final number of fermions inside
the vacuum expectation value, and therefore to reduce as much as
possible the size of the matrix $G$.

The structure of the paper is the following. First we recall some
facts about free fermions, including Wick's theorem. Then we equal
${\bf I}^{(2)}_N(\xi, \zeta, \eta, \mu)$ to a certain vacuum
expectation value, see (\ref{mainequality}). After certain
preliminaries,  via  (\ref{bbbar}), we introduce dressed fermions
$b_{\alpha},{\bar b}_{\alpha}$ to adapt formula
(\ref{mainequality}) for a usage of Wick's theorem in its
determinantal form. Then we consider each of the numbered cases
separately. For the case (1) (which is the most involved case) we
introduce fermionic operators $a_\alpha,{\bar a}_\alpha$, obtained
from $b_{\alpha},{\bar b}_{\alpha}$ via a sort of conjugation (see
(\ref{a1})-(\ref{bar-a2})), and whose vacuum expectation evaluated
via Wick's theorem yields the final result given by formulae
(\ref{mainintegral-ref-1a})-(\ref{02a'-new}) and are illustrated
by six examples. For the case (2), instead of $a_\alpha,{\bar
a}_\alpha$, we similarly use operators
 $b_{\alpha},{\bar b}_{\alpha}$.
 At last, for the case (3), instead of $a_\alpha,{\bar a}_\alpha$, we exploit operators
 $c_\alpha,{\bar c}_\alpha$ introduced in (\ref{c1})-(\ref{bar-c2}).

\br One can obtain the corresponding results for $N$-fold
integrals arising in one-matrix models by specifying that the
measure be proportional to Dirac delta function, say
$\delta(x-y)$. As this specialization will not be consider
detailed here we shall write ${\bf I}_N$ instead of ${\bf
I}^{(2)}_N$. \er


\section{Summary of  free fermions}

  The following is a summary regarding the one and  two-component free fermion algebra based on the
   introductory section of \cite{paper1}. The reader may refer to
  \cite{paper1}, or \cite{JM},  \cite{DJKM} for further details.

\subsection{One component fermions}

In the following, $\AA$ denotes the complex Clifford algebra over
$ \mathbb{C}$ generated by
 \emph{charged free fermions} $\{f_i$, ${\bar f}_i\}_{i\in {\bf
Z} }$, satisfying the anticommutation relations
 \be\label{fermions}
[f_i,f_j]_+=[{\bar f}_i,{\bar f}_j]_+=0,\quad [f_i,{\bar
f}_j]_+=\delta_{ij}.
 \ee
where $[,]_+$ denotes the anticommutator.

Elements of the linear part \be W:=\left(\oplus_{m \in
\Zb}\mathbb{ C}f_m\right)\oplus \left(\oplus_{m\in \Zb}\mathbb{
C}{\bar f}_m\right)
 \ee will be referred to as a {\em free fermions}.  The
 fermionic free fields
\be \label{fermions-fourier}
    f(x):=\sum_{k\in\Zb}f_kx^k,\quad
    {\bar f}(y):=\sum_{k\in\Zb}{\bar f}_ky^{-k-1},
\ee may be viewed as generating functions for the $f_j,
\bar{f}_j$'s.

This Clifford algebra has a standard Fock space representation $F$
and dual space  ${\bar F}$ (see \cite{paper1}) which contain
unique vacuum states $|0\rangle$ and $ \langle 0|$ respectively
satisfying the properties
 \bea \label{vak}
f_m |0\rangle=0 \qquad (m<0),\qquad {\bar f}_m|0\rangle =0 \qquad
(m \ge 0) , \cr
 \langle 0|f_m=0 \qquad (m\ge 0),\qquad \langle
0|{\bar f}_m=0 \qquad (m<0) .
\eea
The {\em Fock spaces} $F$ and
${\bar F}$ are mutually dual, with the hermitian pairing defined
via the linear form $\langle 0| |0 \rangle$ on $\AA$ called the
{\em vacuum expectation value}.  This satisfies
 \bea
\label{psipsi*vac} \langle 0|1|0 \rangle&\&=1;\quad \langle
0|f_m{\bar f}_m |0\rangle=1,\quad m<0; \quad  \langle 0|{\bar
f}_mf_m
|0\rangle=1,\quad m\ge 0 ,\\
\label{end}
 \langle 0| f_n
 |0\rangle&\&=\langle 0|{\bar f}_n
 |0\rangle=\langle 0|f_mf_n |0\rangle=\langle 0|{\bar f}_m{\bar f}_n
 |0\rangle=0;
 \quad \langle 0|f_m{\bar f}_n|0\rangle=0, \quad m\ne n,.
\eea

Wick's theorem implies that for any finite
set of elements $\{w_k \in W\}$, we have
\bea
\label{Wick} \langle 0|w_1
\cdots w_{2n+1}|0 \rangle &\&=0,\cr
 \langle 0|w_1
\cdots w_{2n} |0\rangle &\&=\sum_{\sigma \in S_{2n}} sgn\sigma
\langle 0|w_{\sigma(1)}w_{\sigma(2)}|0\rangle \cdots \langle 0|
w_{\sigma(2n-1)}w_{\sigma(2n)} |0\rangle . \eea Here  $\sigma$
runs over permutations for which $\sigma(1)<\sigma(2),\dots ,
\sigma(2n-1)<\sigma(2n)$ and $\sigma(1)<\sigma(3)<\cdots
<\sigma(2n-1)$.

If  $\{w_i\}_{ i=1,\dots,N}$, are linear combinations of the
$f_j$'s only, $j\in\mathbb{Z}$, and  $\{{\bar w}_i\}_{
i=1,\dots,N}$ linear combinations of the ${\bar f}_j$'s, $j
\in\mathbb{Z}$, then(\ref{Wick}) implies
 \be
 \label{Wick-det}
\langle 0|w_1\cdots w_{N}{\bar w}_N \cdots {\bar w}_1 |0\rangle
=\det\; (\langle 0| w_i{\bar w}_j|0\rangle)\ |_{i,j=1,\dots,N}
 \ee

Following \cite{DJKM},\cite{JM},  for all $ N\in \mathbb{Z}$, we
also introduce the states
 \be \label{1-vacuum}
  \langle  N|:=\langle 0|C_{N}
 \ee
where
 \bea
\label{1-vacuum'} C_{N}&\&:={\bar f}_0\cdots {\bar f}_{N-1}
 \quad {\rm if }\ N>0 \\
C_{N}&\&:={ f}_{-1}\cdots { f}_{N}
\quad {\rm if}\ N<0  \\
 C_{N}&\&:=1 \quad {\rm if}\ N=0
 \eea
and
 \be \label{1-vacuum-r}
    |N \rangle:={\bar C}_{N}|0\rangle
 \ee
where
 \bea
\label{1-vacuum'-r} {\bar C}_{N}&\&:=f_{N-1}\cdots
f_0 \quad {\rm if }\ N>0 \\
{\bar C}_{N}&\&:={\bar f}_{N}\cdots {\bar f}_{-1}
\quad {\rm if}\ N<0  \\
 {\bar C}_{N}&\&:=1 \quad {\rm if}\ N=0
 \eea
The states   (\ref{1-vacuum}) and (\ref{1-vacuum-r}) are referred
to as the left and right charged vacuum vectors, respectively,
with charge $N$.

From the relations
 \be
 \langle 0|  {\bar f}_{N-k} f(x_{i})|0\rangle
 =x_i^{N-k},\quad  \langle 0|
 { f}_{-N+k-1}  {\bar f}(y_{i})|0\rangle =y_i^{N-k},\quad k=1,2,\dots N,
 \ee
 and  (\ref{Wick-det}), it follows that
\bea\label{Delta-N-left} \langle N|f(x_n)\cdots
f(x_1)|0\rangle  &\&=\delta_{n,N}\Delta_N(x),\quad N\in \mathbb{Z},\\
\label{Delta-N-right} \langle -N|\bar{f}(y_n)\cdots
\bar{f}(y_1)|0\rangle&\&=\delta_{n,N}\Delta_N(y),\quad N\in
\mathbb{Z}. \eea

For free fermion generators with $|x|\ne|y|$,
 \be\label{f-barf}
    \langle 0|f(x){\bar f}(y)|0 \rangle
    =    \frac{1}{x-y}
 \ee
Note that the expression on the right hand side is
 actually defined, by (\ref{psipsi*vac}), as the infinite series
 $\sum_{n=0}^\infty y^nx^{-n-1}$
 which converges only inside $|x|<|y|$.
 However one can consider expression (\ref{f-barf}) for the whole
 region of $x$ and $y$ (when $|x|\ne|y|$) in the sense of
 analytical continuation.

From Wick's theorem it follows that
 \be\label{multi-f-bar-f-n-m}
    \langle  n-m|f(x_n)\cdots f(x_1){\bar f}(y_m)\cdots {\bar f}(y_1)|0 \rangle
=   \frac{\Delta_n(x)\Delta_m(y)}
    {\prod_{i=1,\dots,n\atop j=1,\dots,m}(x_i-y_j)}
 \ee

\subsection{Two-component fermions}

The two-component fermion formalism is obtained by relabelling
the above as follows.
 \bea
  \label{2-fermions}
f_n^{(\alpha)}&\&:=f_{2n+\alpha-1}\ ,\qquad \qquad {\bar
f}_n^{(\alpha)}:={\bar f}_{2n+\alpha-1}\ ,
\\
 \label{2-fermions-z}
f^{(\alpha)}(z)&\&:=\sum_{k=-\infty}^{+\infty}z^kf_{k}^{(\alpha)}\
,\quad {\bar
f}^{(\alpha)}(z):=\sum_{k=-\infty}^{+\infty}z^{-k-1}{\bar
f}_{k}^{(\alpha)}\ ,
 \eea
where $\alpha=1,2$. Then (\ref{fermions}) is equivalent to
 \be\label{2-fermions-antic}
[f_n^{(\alpha)},f_m^{(\beta)}]_+=[{\bar f}_n^{(\alpha)},{\bar
f}_m^{(\beta)}]_+=0,\qquad [f_n^{(\alpha)},{\bar
f}_m^{(\beta)}]_+=\delta_{\alpha,\beta}\delta_{nm}.
 \ee

We denote the right and left vacuum vectors respectively as \be
\label{2-vacuum-def} |0,0\rangle:=|0\rangle ,\quad \langle
0,0|:=\langle 0| . \ee Relations (\ref{vak}) then become, for
$\alpha=1,2$,
\begin{eqnarray}\label{2-vak-r}
f_m^{(\alpha)} |0,0\rangle=0 \qquad (m<0),\qquad {\bar
f}_m^{(\alpha)}|0,0\rangle =0 \qquad (m \ge 0) , \\
\label{2-vak-l} \langle 0,0|f_m^{(\alpha)}=0 \qquad (m\ge
0),\qquad \langle 0,0|{\bar f}_m^{(\alpha)}=0 \qquad (m<0) .
\end{eqnarray}

For $n,m\in\mathbb{Z},\ i,j=1,2$ it follows from
(\ref{psipsi*vac})-(\ref{end}) that
 \be
\label{00ff00-1}
 \langle  0,0|{ f}_n^{(i)}{ f}_m^{(j)}|0,0\rangle=
 \langle  0,0|{\bar f}_n^{(i)}{\bar f}_m^{(j)}|0,0\rangle=0,
 \ee
 \be
\label{00ff00-2} \langle  0,0|{ f}_n^{(i)}{\bar
f}_m^{(j)}|0,0\rangle=\delta_{ij}\delta_{nm},\ n<0
 \ee

 Following \cite{DJKM},\cite{JM}, we also introduce the states
 \be \label{2-vacuum}
  \langle  n^{(1)},n^{(2)}|:=\langle 0,0|C_{n^{(1)}}C_{n^{(2)}},
 \ee
where \bea \label{2-vacuum'} C_{n^{(\alpha)}}&\&:={\bar
f}^{(\alpha)}_0\cdots {\bar f}^{(\alpha)}_{n^{(\alpha)}-1}
 \quad {\rm if }\ n^{(\alpha)}>0 \\
 \label{2-vacuum'-neg}
C_{n^{(\alpha)}}&\&:={ f}^{(\alpha)}_{-1}\cdots {
f}^{(\alpha)}_{n^{(\alpha)}}
 \qquad {\rm if}\ n^{(\alpha)}<0  \\
 C_{n^{(\alpha)}}&\&:=1 {\hskip 79 pt} {\rm if}\ n^{(\alpha)}=0
 \eea
and
 \be \label{2-vacuum-r}
    |n^{(1)},n^{(2)}\rangle:={\bar C}_{n^{(2)}}{\bar C}_{n^{(1)}}|0,0\rangle
 \ee
where
 \bea
\label{2-vacuum'-r} {\bar
C}_{n^{(\alpha)}}&\&:=f^{(\alpha)}_{n^{(\alpha)}-1}\cdots
f^{(\alpha)}_0 \quad {\rm if }\ n^{(\alpha)}>0 \\
 \label{2-vacuum'-r-neg}
{\bar C}_{n^{(\alpha)}}&\&:={\bar
f}^{(\alpha)}_{n^{(\alpha)}}\cdots {\bar f}^{(\alpha)}_{-1}
\qquad {\rm if}\ n^{(\alpha)}<0  \\
 {\bar C}_{n^{(\alpha)}}&\&:=1  {\hskip 79 pt}  {\rm if}\ n^{(\alpha)}=0
 \eea
The states (\ref{2-vacuum}) and (\ref{2-vacuum-r}) will be
referred to, respectively, as left and right charged vacuum
vectors with charges $(n^{(1)},n^{(2)})$.

It follows that
\begin{eqnarray}\label{2-vak-ch-1-r}
f_m^{(1)} |n,*\rangle=0 \qquad (m<n),\qquad {\bar
f}_m^{(1)}|n,*\rangle =0 \qquad (m \ge n) , \\
\label{2-vak-ch-1-l} \langle n,*|f_m^{(1)}=0 \qquad (m\ge
n),\qquad \langle n,*|{\bar f}_m^{(1)}=0 \qquad (m<n) ,
\end{eqnarray}
and
\begin{eqnarray}\label{2-vak-ch-2-r}
f_m^{(2)} |*,n\rangle=0 \qquad (m<n),\qquad {\bar
f}_m^{(2)}|*,n\rangle =0 \qquad (m \ge n) , \\
\label{2-vak-ch-2-l} \langle *,n|f_m^{(2)}=0 \qquad (m\ge
n),\qquad \langle *,n|{\bar f}_m^{(2)}=0 \qquad (m<n) .
\end{eqnarray}

\br \label{shifts} Note that if we shift the charges of the vacuum vectors
 \be\label{shift1}
\langle *,*|\to\langle  *,*+n|,\quad |*,*\rangle \to |*,*+n\rangle
  \ee
  and, at the same time, re-label
 \be\label{shift2}
f^{(2)}_{i}\to f^{(2)}_{i+n},\quad {\bar f}^{(2)}_{i}\to {\bar
f}^{(2)}_{i+n},\quad i,n\in\mathbb{Z},
 \ee
 then the vacuum expectation values remain invariant. \er

\br \label{2Wick}
Wick's theorem will be used in the form
(\ref{Wick-det}) in the following. In the two component notation, there are two
possible ways to do this; either

(1) Use formula  (\ref{Wick-det}), remembering that 2-component
fermions consist just  of the usual even and odd ones  (see
(\ref{2-fermions})), or

(2) Use formula (\ref{Wick-det}) separately for each component. To
calculate the vacuum state expectation value of an operator $O$,
we first present it in the form
 \be\label{O-comp-decomp}
O=\sum_i O^{(1)}_i O^{(2)}_i.
 \ee
Then
 \be
 \label{Wick-comp-decomp}
\langle 0,0|O|0,0\rangle=\sum_i \langle 0,0|O^{(1)}_i O^{(2)}_i
|0,0\rangle=\sum_i \langle 0,0|O^{(1)}_i|0,0\rangle \langle
0,0|O^{(2)}_i |0,0\rangle
 \ee
where Wick's theorem in the form (\ref{Wick-det}) may be applied
to each factor $\langle 0,0|O^{(\alpha)}_i|0,0\rangle$

The version (2) is used in the Section 3, and the version (1) is
used in the Section 5. \er

Now define \bea F^{(j)}(x^{(j)})&\&:=f^{(j)}(x_{n_j}^{(j)})\cdots
f^{(j)}(x_1^{(j)}), \cr {\bar F}^{(j)}(y^{(j)})&\&:={\bar
f}^{(j)}(y_{m_j}^{(j)})\cdots {\bar f}^{(j)}(y_1^{(j)}),\quad
j=1,2 \eea

Combining (\ref{multi-f-bar-f-n-m}) and (\ref{Wick-comp-decomp})
we obtain
\bea
&\&    \langle  n_1-m_1,n_2-m_2|F^{(2)}(x^{(2)})F^{(1)}(x^{(1)})
     {\bar F}^{(1)}(y^{(1)}){\bar F}^{(2)}(y^{(2)})|0,0\rangle \cr
&\&
 =   (-1)^{m_2(m_1+n_1)} \frac{\Delta_{n_1}(x^{(1)})\Delta_{m_1}(y^{(1)})
 \Delta_{n_2}(x^{(2)})\Delta_{m_2}(y^{(2)})}
    {\prod_{i=1,\dots,n_1\atop j=1,\dots,m_1}(x_i^{(1)}-y_j^{(1)})
    \prod_{i=1,\dots,n_2\atop j=1,\dots,m_2}(x_i^{(2)}-y_j^{(2)})}.
    \label{multi-f-bar-f-n-m-2-comp}
 \eea

expectation value 

\section{The integral of rational symmetric functions as a certain
expectation value}

We shall use the following notations
 \be
\eta=(\eta_1,\dots,\eta_{M_1}),\quad \xi=(\xi_1,\dots,\xi_{L_1}),
\quad \mu=(\mu_1,\dots,\mu_{M_2}),\quad
\zeta=(\zeta_1,\dots,\zeta_{L_2}) \ee \be N_j=N+L_j-M_j, \quad
j=1,2, \ee
 \be
\int d\mu({\bf x},{\bf y})(\cdot )={\int\int} d\mu(x_1,y_1) \dots
{\int\int} d\mu(x_N,y_N)(\cdot )
 \ee

Let
 \be\label{F1}
 {
F}^{(1)}(\xi) :=f^{(1)}(\xi_{L_1})\cdots f^{(1)}(\xi_1),
 \ee
 \be\label{}
{ F}^{(2)}(\mu):={ f}^{(2)}(\mu_{M_2})\cdots { f}^{(2)}(\mu_1),
 \ee
 \be\label{}
{\bar F}^{(1)}(\eta):={\bar f}^{(1)}(\eta_{M_1})\cdots {\bar
f}^{(1)}(\eta_1),
 \ee
 \be\label{F2bar}
{\bar F}^{(2)}(\zeta):={\bar f}^{(2)}(\zeta_{L_2})\cdots {\bar
f}^{(2)}(\zeta_1),
 \ee
and
 \be g:=e^{A}, \quad A:={\int\int} f^{(1)}(x){\bar
f}^{(2)}(y)d\mu(x,y), \label{A-for-2MM} \ee and consider the
expression
 \be\label{M-1-L1-M-2-L2-insertions}
J_N(\xi, \zeta, \eta, \mu) :=R_N \langle N_1,-N_2| { F}^{(2)}(\mu)
F^{(1)}(\xi)\; g\; {\bar F}^{(1)}(\eta){\bar F}^{(2)}(\zeta)
|0,0\rangle
 \ee
where
 \be
 \label{R_N}
R_N=R_N(\xi, \zeta, \eta, \mu):={s(L_1,L_2,M_1,M_2)\over
\prod_{n=0}^{N-1}h_n}\frac {
\prod_{\alpha=1}^{L_1}\prod_{j=1}^{M_1}(\xi_\alpha -
\eta_j)\prod_{\beta=1}^{L_2}\prod_{k=1}^{M_2}(\zeta_\beta-\mu_k)
 }
 {\Delta_{L_1}(\xi)\Delta_{L_2}(\zeta)\Delta_{M_1}(\eta)\Delta_{M_2}(\mu)},
 \ee
and $s(L_1,L_2,M_1,M_2)$ is the sign factor \be
s(L_1,L_2,M_1,M_2))=(-1)^{{1\over 2}
N(N+1)+L_2(L_1+M_1)+N(L_1+M_1)+M_2L_2}  \ee

\br \label{on-g} Because of the form (\ref{A-for-2MM}), $g$
commutes with both  ${ F}^{(1)}(\xi)$ and ${\bar F}^{(2)}(\mu)$.
Moreover he conditions given in Remark (\ref{rem-conditions}) also
imply that $g$ commutes with ${\bar F}^{(1)}(\eta)$ and ${
F}^{(2)}(\zeta)$.
 \er

The main result  of this subsection is the equality
 \be
{\bf I}_N(\xi, \zeta, \eta, \mu)=J_N(\xi, \zeta, \eta, \mu)
 \label{mainequality}
 \ee

\noindent
 {\bf Proof.}
Inserting
\[
g=\sum_{n=0}^\infty \frac{A^n}{n!}
\]
 into
(\ref{M-1-L1-M-2-L2-insertions}), we note that only the $N$-th power
contributes, giving
 \be
J_N(\xi, \zeta, \eta, \mu) =\frac{R_N}{N!}\langle N_1,-N_2|{
F}^{(2)}(\mu) F^{(1)}(\xi)A^N{\bar F}^{(1)}(\eta) {\bar
F}^{(2)}(\zeta) |0,0\rangle \label{L1-L2-insertions} \ee
Collecting the fermion terms of the same types (which gives rise
to a sign factor), the right hand side of (\ref{L1-L2-insertions})
may be expressed as
 \[
(-1)^{\frac12N(N-1) + NM_1}\frac{R_N}{N!} \int d\mu({\bf x},{\bf
y})
 \langle N_1,-N_2|{
F}^{(2)}(\mu) F^{(1)}(\xi,x){\bar F}^{(1)}(\eta) {\bar
F}^{(2)}(y,\zeta) |0,0\rangle
 \]
 where $(\xi,x)=(\xi_1,\dots,\xi_{L_1},x_1,\dots,x_N)$ and
$(y,\zeta)=(y_1,\dots,y_N,\zeta_1,\dots,\zeta_{L_2})$. Using
(\ref{multi-f-bar-f-n-m-2-comp}) to  evaluate the expectation
value in the integrand we get

\[
 J_N(\xi, \zeta, \eta, \mu)
={1\over  {\bf Z}^{(2)}_N} \int d\mu({\bf x},{\bf y})
 \Delta_N(x)\Delta_N(y)
 \prod_{a=1}^N {\prod_{\alpha=1}^{L_1}(\xi_\alpha - x_a)
 \prod_{\beta=1}^{L_2}(\zeta_\alpha - y_a)
 \over \prod_{j=1}^{M_1}(\eta_j - x_a)  \prod_{k=1}^{M_2}(\mu_k - y_a)}
 \]
 \[
= {\bf I}_N(\xi, \zeta, \eta, \mu) .
 \]
 \br From (\ref{M-1-L1-M-2-L2-insertions}) and (\ref{mainequality})
 in the absence of the $F^{(1)}, F^{(2)}, \bar{F}^{(1)}, \bar{F}^{(2)}$
 terms,  we obtain the representation which we used in
 \cite{paper1}
\[
 {\bf Z}^{(2)}_N=(-1)^{\frac 12 N(N+1)}N!\langle N,-N| g|0,0\rangle.
\]
 \er

\;

 Having the representation (\ref{mainequality}) we evaluate
 ${\bf I}_N(\xi, \zeta, \eta, \mu) $ via
 Wick's theorem. Before we need certain preliminary relations.


\section{Biorthogonal polynomials and dressed fermions}

In this section we introduce transformations $\Omega$ and $Q$ and
dressed fermions $d^{(\alpha)},{\bar d}^{(\alpha)}$ and
$b_{\alpha},{\bar b}_{\alpha}$. We re-write the v.e.v.
(\ref{M-1-L1-M-2-L2-insertions}) in a
 form suitable for further calculation by means of Wick's theorem.

(a) {\bf Biorthogonal polynomials}.

Consider the sequence  of biorthogonal polynomials associated to
the measure $d\mu(x,y)$:
 \be\label{orth-rel}
 {\int\int} P_n(x)S_m(y)d\mu(x,y)=\delta_{n,m},\quad n,m\ge 0
 \ee

 It is convenient to
 write down the biorthogonal polynomials in the following form
 \be\label{K-p-K-s}
P_n(x)=\frac{1}{\sqrt{h_n}}\sum_{m= 0}^n K_{nm}x^m ,\quad
S_n(y)=\frac{1}{\sqrt{h_n}}\sum_{ m= 0}^n y^m({\bar K}^{-1})_{mn}
,\quad n\ge 0
 \ee
 where $K_{nm}$ and $({\bar K}^{-1})_{mn}$ are respectively viewed as entries
 of semi-infinite matrices $K$ and  $({\bar K}^{-1})$. $K$ is a lower triangular
 matrix and ${\bar K}$ is an upper triangular one, both have units
 on the diagonal: $K_{nn}={\bar K}_{nn}=1$, $n=0,1,2,\dots$.

  As is well-known, the orthogonality relation (\ref{orth-rel})
implies
 \be\label{factorization}
H{\bar K}=KB,
 \ee
where $H=diag(h_n)$ and $B$ is the bi-moment matrix,
 \be
\label{bimoments} B_{nm}= \int\int x^ny^{m}d\mu(x,y) ,\quad m,n\ge
0,
 \ee

In time (see \cite{GMO}), (\ref{factorization}) was made good use
to relate matrix models to integrable systems via the
factorization method widely used in soliton theory starting from
the basic paper \cite{ZSh}.
 Then $K$ and ${\bar
K}$ may be identified with the Mikhailov-Ueno-Takasaki dressing
matrices (see \cite{AM},\cite{UT}), whose rows give rise to a pair
the so-called Baker functions, while columns of $K^{-1}$and of
${\bar K}^{-1}$ give rise to a pair of adjoint Baker functions. In
case the dressing matrices are semi-infinite, the first of the
Baker functions (related to rows of $K$) and the second of the
adjoint Baker function (related to columns ${\bar K}^{-1}$) take
the form of quasi-polynomials.

Here we shall use (\ref{factorization}) differently.

\;

(b) {\bf  Canonical transformation $\Omega$}.

First, we introduce
 \be\label{K-p*-K-s*}
\tilde{S}_n(x):=\sum_{m= 0}^{+\infty}
x^{-m-1}(K^{-1})_{mn}\sqrt{h_n} ,\quad \tilde{P}_n(y):=\sum_{ m=
0}^{+\infty} {\bar K}_{nm}y^{-m-1}\sqrt{h_n},\quad n\ge 0
 \ee

 Properties of $\tilde{S}_n(x)$ and $\tilde{P}_n(y)$ are described
 in the Appendix C. The main is that they coincide with Hilbert
 transforms of biorthogonal polynomials (\ref{HT}).

Now, let us define matrices $\omega^{(1)}$ and $\omega^{(2)}$ via
 \be\label{omega-K}
 e^{\omega^{(1)}}=K,\quad e^{\omega^{(2)}}={\bar K}
 \ee
(Respectively, strictly upper and strictly lower) triangular
matrices $\omega^{(1),(2)}$ may be defined in a unique way by a
recursion procedure.

We introduce
  \be\label{Omega}   \Omega=\exp
\sum_{n>m\ge 0} \left(\omega_{n,m}^{(1)}f_n^{(1)}{\bar
f}_{m}^{(1)} + \omega_{m,n}^{(2)}f_{-m-1}^{(2)}{\bar
f}_{-n-1}^{(2)}\right)
 \ee

Note that, for any $N_1$ and $N_2$,
 \be\label{Omega-vac-j}
\langle N_1,-N_2|{ \Omega}^{-1}=\langle N_1,-N_2|\ ,\quad {
\Omega}|0,0\rangle =|0,0\rangle
 \ee
The first equality in (\ref{Omega-vac-j}) follows from
(\ref{2-vak-ch-1-l}),(\ref{2-vak-ch-2-l}) and from the restriction
$n>m$ in the summation in (\ref{Omega}). The second equality in
(\ref{Omega-vac-j}) follows from the restriction $m\ge 0$ and from
(\ref{2-vak-ch-1-l}),(\ref{2-vak-ch-2-l}).

 Consider
  \be\label{d1''} d^{(\alpha)}(x)={
\Omega}f^{(\alpha)}(x){ \Omega}^{-1},\quad {\bar
d}^{(\alpha)}(x)={ \Omega}{\bar f}^{(\alpha)}(x){
\Omega}^{-1},\quad \alpha=1,2
 \ee
 which we refer as dressed fermions.

We have (see Appendix B)
 \be\label{d-f-p}
d^{(1)}(x) =\sum_{n=-\infty}^{+\infty} f_n^{(1)}P_n(x)\sqrt{h_n}
 \ee
 \be\label{d-f-p*}
{\bar d}^{(1)}(x) = \sum_{n=-\infty}^{+\infty}  {\bar
f}_n^{(1)}{\tilde {S}_n(x)\over \sqrt{h_n}}
 \ee
 \be\label{d-f-s*}
{ d}^{(2)}(y) =\sum_{n=-\infty}^{+\infty}  {
f}_{-n-1}^{(2)}{\tilde {P}_n(y)\over \sqrt{h_n}}
 \ee
 \be\label{d-f-s}
{\bar d}^{(2)}(y) =\sum_{n=-\infty}^{+\infty}  {\bar
f}_{-n-1}^{(2)}S_n(y)\sqrt{h_n}
 \ee
where for $n<0$ we use the following notational convention
 \be\label{PS-neg}
 P_n(x)=S_n(x)= x^n  ,\quad \tilde
 {S}_n(x)=\tilde {P}_n(x)=x^{-n-1}, \quad h_n=1,\quad n<0
 \ee

 A kind of dressed fermionic operators (where powers of $x$ were replaced
 by Baker functions similar to (\ref{d-f-p})-(\ref{d-f-s})),
 in a different context,  were also introduced in  \cite{GO}
where they were called Krichever-Novikov fermions (see Appendix to
\cite{GO}).

If we write
 \[
d^{(\alpha)}(x) =\sum_{n=-\infty}^{+\infty} d_n^{(\alpha)}x^n ,
 \quad
{\bar d}^{(\alpha)}(x) = \sum_{n=-\infty}^{+\infty}  {\bar
d}_n^{(\alpha)}x^{-n-1},\quad \alpha=1,2
 \]
 where
 \[
d_n^{(\alpha)}={ \Omega}f^{(\alpha)}_n{ \Omega}^{-1} ,\quad {\bar
d}_n^{(\alpha)}={ \Omega}{\bar f}^{(\alpha)}_n{ \Omega}^{-1},
 \]
 then
 \be\label{d1}
 d_i^{(1)}=
\sum_{n\ge i\ge 0}f_n^{(1)}K_{ni}, \quad i\ge 0, \qquad d_i^{(1)}=
{f}_{i}^{(1)},\quad i < 0
 \ee
 \be\label{d1bar}
{\bar d}_i^{(1)}= \sum_{i\ge n\ge 0}(K^{-1})_{in}{\bar f}_n^{(1)},
\quad i\ge 0, \qquad {\bar d}_i^{(1)}= {\bar f}_{i}^{(1)},\quad i
< 0
 \ee
 \be\label{d2}  {
d}_{-i-1}^{(2)}= \sum_{i\ge n\ge 0}{ f}_{-n-1}^{(2)}{\bar K}_{ni},
\quad i\ge 0, \qquad { d}_{-i-1}^{(2)}={f}_{-i-1}^{(2)},\quad i <
0
 \ee
 \be\label{d2bar}
 {\bar d}_{-i-1}^{(2)}=
\sum_{n\ge i\ge 0}({\bar K}^{-1})_{in}{\bar f}_{-n-1}^{(2)}, \quad
i\ge 0, \qquad {\bar d}_{-i-1}^{(2)}={\bar f}_{-i-1}^{(2)},\quad i
< 0
 \ee

\;

(c) {\bf Useful formulae for  charged vacuum vectors}.

 First, we introduce
 \be\label{}
g_n := e^{f^{(1)}_n{\bar f}^{(2)}_{-n-1}} =1+f^{(1)}_n{\bar
f}^{(2)}_{-n-1} ,
 \ee
 \be\label{}
e_n := e^{{\bar f}^{(1)}_n f^{(2)}_{-n-1}} =1+ {\bar f}^{(1)}_n
f^{(2)}_{-n-1}
 \ee
We see that $g_n,e_n\in {\hat{GL}}_\infty $.

We shall also consider powers of these operators
 \be\label{g-n-h}
(g_n)^{p} =e^{p f^{(1)}_n{\bar f}^{(2)}_{-n-1}} =1+p
f^{(1)}_nf^{*(2)}_{-n-1},
 \ee
 \be\label{e-n-h}
(e_n)^{p} =e^{p {\bar f}^{(1)}_n f^{(2)}_{-n-1}} =1+p {\bar
f}^{(1)}_n f^{(2)}_{-n-1}
 \ee
where $n\in\mathbb{Z},p\in\mathbb{C}$.

\;

 Now,  due to
 \[
\langle 0,0|e_n g_n =\langle 0,0|(1+ {\bar f}^{(1)}_n
f^{(2)}_{-n-1})(1+f^{(1)}_n{\bar f}^{(2)}_{-n-1})=\langle
0,0|(1-1+{\bar f}^{(1)}_n f^{(2)}_{-n-1})=\langle 0,0|{\bar
f}^{(1)}_n f^{(2)}_{-n-1} ,
 \]
 which is true for $n>0$,
and due to (\ref{g-e-1}), (\ref{g-e-2}), we conclude that for
$N>0$
 \[
\langle N,-N|=(-1)^{\frac12 N(N-1)}\langle
0,0|\prod_{n=0}^{N-1}e_n\prod_{n=0}^{N-1}g_n
 \]
where  $\langle  N,- N|$ was defined in
(\ref{2-vacuum})-(\ref{2-vacuum'-neg}).

In the similar way we obtain a representation we shall need later
 \be\label{vac-G-1}
\quad N_2> N_1\ge 0:\quad \langle N_1,-N_2|=(-1)^{\frac12
N_1(N_1-1)}\langle 0,0|f^{(2)}_{-N_1-1}\cdots
f^{(2)}_{-N_2}\prod_{n=0}^{N_1-1}e_n\prod_{n=0}^{N_1-1}g_n
 \ee

\;

(d) {\bf Evaluation of $\Omega g|0,0\rangle $}.

 Taking into account that $f_{-n-1}^{(1)}|0,0\rangle ={\bar f}_{n}^{(2)}|0,0\rangle=0$
 for $n>0$, one obtains
 \be\label{}
\Omega g|0,0\rangle=e^{\sum_{i,j,n,m\ge 0 }K_{in}B_{nm}{\bar
K}_{mj}f_i^{(1)}{\bar f}_{-j-1}^{(2)}}|0,0\rangle
 =  Q^{-1}|0,0\rangle,\quad
Q^{-1}:=\prod_{n=0}^{+\infty}(g_n)^{h_n}
 \ee
 where we used (\ref{d1}),(\ref{d2bar}) and (\ref{factorization}).

\;

(e) {\bf Fermionic operators $b_\alpha,{\bar b}_\alpha$}.

We shall also need the following fermionic operators
 \be\label{bbbar} b_\alpha(x):=
Qd^{(\alpha)}(x)Q^{-1}=Q\Omega \; f^{(\alpha)}(Q\Omega)^{-1},\quad
{\bar b}_\alpha(x):= Q{\bar d}^{(\alpha)}(x)Q^{-1}=Q\Omega \;
{\bar f}^{(\alpha)}(Q\Omega)^{-1}
 \ee

Using (\ref{g-e-f-1})-(\ref{e-f-3'}) we write down
 \be\label{b1}
b_1(\xi)=
\sum_{n=-\infty}^{+\infty}{f}_{n}^{(1)}P_n(\xi)\sqrt{h}_n
=d^{(1)}(\xi) \ ,
 \ee
 \be\label{bar-b1}
{\bar b}_1(\eta)=  \sum_{n=-\infty}^{+\infty}{\bar
f}_n^{(1)}{\tilde{S}_n(\eta)\over
\sqrt{h}_n}+\sum_{n=0}^{+\infty}{\bar
f}_{-n-1}^{(2)}{\tilde{S}_n(\eta) \sqrt{h}_n} \ ,
 \ee
 \be\label{b2}
b_2(\mu)=
\sum_{n=-\infty}^{+\infty}f_{-n-1}^{(2)}{\tilde{P}_n(\mu)\over
\sqrt{h}_n} -\sum_{n=0}^{+\infty}f_{n}^{(1)}{\tilde{P}_n(\mu)
\sqrt{h}_n} \ ,
 \ee
 \be\label{bar-b2}
{\bar b}_2(\zeta)= \sum_{n=-\infty}^{+\infty}{\bar
f}_{-n-1}^{(2)}S_n(\zeta)\sqrt{h}_n={\bar d}^{(2)}(\zeta)
 \ee

 Notice that each of ${\bar b}_1$ and $b_2$ contains both component
 fermions.

The fermionic operators $b_\alpha,{\bar b}_\alpha $ may be also
called dressed fermions. Similarly to (\ref{F1})-(\ref{F2bar}),
for their products, we shall use large letters, namely, we define
\[
B_1(\xi):= Q\Omega\; F^{(1)}(\xi)\;(Q\Omega)^{-1},\quad B_2(\mu):=
Q\Omega \; F^{(2)}(\mu)\;(Q\Omega)^{-1},
\]
\[
{\bar B}_1(\eta):= Q\Omega\;{\bar F}^{(1)}(\eta)\;(Q\Omega)^{-1}
,\quad {\bar B}_2(\zeta):= Q\Omega\;{\bar
F}^{(2)}(\zeta)\;(Q\Omega)^{-1}
\]

\;

(f) {\bf ${\bf I}_N$ as a re-written vacuum expectation value}

Now we restate (\ref{mainequality}) as
 \be\label{mainequality-B}
{\bf I}_N(\xi, \zeta, \eta, \mu)R_N^{-1} =
 \langle N_1,-N_2|Q^{-1}\; {
B}_2(\mu) B_1(\xi) {\bar B}_1(\eta){\bar B}_2(\zeta) |0,0\rangle
 \ee
where $R_N$ was defined by (\ref{R_N}).

Now we have to consider the cases (1),(2),(3) separately.


\section{Determinantal expression for integrals of rational symmetric functions}

Below we apply Wick's theorem to evaluate v.e.v. in the right hand
side of  (\ref{mainequality-B}), getting answers  for ${\bf
I}^{(2)}_N(\xi, \zeta, \eta, \mu)$ for all the cases listed in the
Introduction.


\subsection{Determinantal representation in the case (1): $N_2\ge N_1\ge 0$}

In the case (1)
 we have the following formulae
for $\langle N_1,-N_2|Q^{-1} $:
 \be\label{vacE'}
\ N_2\ge N_1\ge 0:\quad\langle N_1,-N_2|Q^{-1} =(-1)^{\frac12
N_1(N_1+1)}\langle 0,0|f^{(2)}_{-N_1-1}\cdots f^{(2)}_{-N_2}E
\prod_{n=0}^{N_1-1}h_n \ ,
 \ee
where
 \be\label{E,S}
E:=\prod_{n=0}^{N_1-1}(e_n)^{-h_n^{-1}}
 \ee

Formula (\ref{vacE'}) follows from relations
\[
\langle 0,0|e_ng_n^{1+h_n}=\langle 0,0|(1+{\bar f}_n^{(1)}{
f}_{-n-1}^{(2)})(1+(1+h_n){ f}_n^{(1)}{\bar f}_{-n-1}^{(2)})=
\langle 0,0|(-h_n+{\bar f}_n^{(1)}{ f}_{-n-1}^{(2)})
\]
\[
=-h_n\langle 0,0|e_n^{-h_N^{-1}},\quad n\ge 0
\]
and from (\ref{vac-G-1}).

It is important that
 \be\label{Evac}
E|0,0\rangle=0 \ ,
 \ee
which is true as each $e_n^p|0,0\rangle=0,n\ge 0$.

 Using (\ref{g-e-f-1})-(\ref{e-f-3'}), we may
evaluate
 \be\label{a1}
a_1(\xi):= E b_{1}(\xi) E^{-1} =
\sum_{n=-\infty}^{+\infty}f_n^{(1)}P_n(\xi)\sqrt{h_n}+
\sum_{n=0}^{N_1-1}f_{-n-1}^{(2)}{P_n(\xi)\over\sqrt{h_n}} \ ,
 \ee
 \be\label{bar-a1}
{\bar a}_1(\eta):= E{\bar b}_{1}(\eta)E^{-1}  =
\sum_{n=N_1}^{+\infty}{\bar f}_n^{(1)}{\tilde {S}_n(\eta)\over
\sqrt{h}_n}+ \sum_{n=-\infty}^{-1}{\bar f}_n^{(1)}{\tilde
{S}_n(\eta)\over \sqrt{h}_n}+ \sum_{n=0}^{+\infty}{\bar
f}_{-n-1}^{(2)}{\tilde {S}_n(\eta) \sqrt{h}_n} \ ,
 \ee
 \be\label{a2}
a_2(\mu):=  E b^{(2)}(\mu)E^{-1}=
\sum_{n=N_1}^{+\infty}f_{-n-1}^{(2)}{\tilde{P}_n(\mu)\over\sqrt{h}_n}
+
\sum_{n=-\infty}^{-1}f_{-n-1}^{(2)}{\tilde{P}_n(\mu)\over\sqrt{h}_n}
-\sum_{n=0}^{+\infty}f_{n}^{(1)}{\tilde{P}_n(\mu)\sqrt{h}_n} \ ,
 \ee
 \be\label{bar-a2}
{\bar a}_2(\zeta):=  E {\bar b}^{(2)}(\zeta)E^{-1} =
\sum_{n=-\infty}^{+\infty}{\bar
f}_{-n-1}^{(2)}S_n(\zeta)\sqrt{h_n}- \sum_{n=0}^{N_1-1}{\bar
f}_{n}^{(1)}{S_n(\zeta)\over\sqrt{h_n}}
 \ee

Using notations
\[
A_1(\xi):= EQ\Omega\; F^{(1)}(\xi)\;(EQ\Omega)^{-1},\quad
A_2(\mu):= EQ\Omega \; F^{(2)}(\mu)\;(EQ\Omega)^{-1},
\]
\[
{\bar A}_1(\eta):= EQ\Omega\;{\bar F}^{(1)}(\eta)\;(EQ\Omega)^{-1}
,\quad {\bar A}_2(\zeta):= EQ\Omega\;{\bar
F}^{(2)}(\zeta)\;(EQ\Omega)^{-1}
\]
for the products, by analogy with (\ref{F1})-(\ref{F2bar}), by
(\ref{vacE'}) and (\ref{Evac}), we  arrive at
 \be\label{step3}
 {\bf I}_N(\xi, \zeta, \eta, \mu)R_N^{-1}
  =
(-1)^{\frac12 N_1(N_1+1)}\prod_{n=0}^{N_1-1}h_n\langle
0,0|f^{(2)}_{-N_1-1}\cdots f^{(2)}_{-N_2} { A}_2(\mu)A_1(\xi)
{\bar A}_1(\eta){\bar A}_2(\zeta) |0,0\rangle
 \ee

Finely, this form is applicable to apply the Wick's formula
(\ref{Wick-det}). Indeed, each $a_i$ is a linear combination of
fermions $f^{(1)}$ and $f^{(2)}$, while each ${\bar a}_i$ is a
linear combination of fermions ${\bar f}^{(1)}$ and ${\bar
f}^{(2)}$.

Then, by (\ref{Wick-det}), the vacuum expectation value in formula
(\ref{step3}) is equal to the determinant  of a $L_1+M_2$ by
$L_1+M_2$  matrix, which consists of six blocks formed by
pair-wise v.e.v.:

$$
\left(\begin{array}{ccccc}
 \langle { a}_1(\xi_\alpha){\bar a}_1(\eta_j)\rangle
 &\langle{ a}_2(\mu_k){\bar a}_1 (\eta_j)\rangle
  & \langle f^{(2)}_{i-1-N-L_2+M_2}{\bar a}_1(\eta_j)\rangle \\
\langle{ a}_1(\xi_\alpha){\bar a}_2 (\zeta_\beta)\rangle
 & \langle{ a}_2(\mu_k){\bar a}_2 (\zeta_\beta)\rangle
   & \langle f^{(2)}_{i-1-N-L_2+M_2}{\bar a}_2(\zeta_\beta)\rangle\\
\end{array}\right)
$$
where
\[
\alpha=1,\dots,L_1;\quad \beta=1,\dots,L_2;\quad   j= 1,\dots,M_1
;\quad k=1,\dots,M_2
\]
\[
i=1,\dots,M_1+L_2-L_1-M_2
\]

Now, taking into account (\ref{00ff00-2})-(\ref{00ff00-2}), from
the  explicit formulae (\ref{a1})-(\ref{bar-a2}), and also from
(\ref{sum-ss*}) and (\ref{sum-s*p*}), we compute all relevant
pair-wise vacuum expectation values
 \[
\langle { a}_1(\xi_i){\bar
a}_1(\eta_j)\rangle=\frac{1}{\xi_i-{\eta_j}}+\sum_{n=0}^{N_1-1}P_n(\xi_i)\tilde{S}_n(\eta_j),
 \]
 \[
\langle { a}_2(\mu_i){\bar
a}_2(\zeta_j)\rangle=\sum_{n=N_1}^{+\infty}S_n(\zeta_j)\tilde{P}_n(\mu_i)=-\frac{1}{\zeta_j-\mu_i}-
\sum_{n=0}^{N_1-1}S_n(\zeta_j)\tilde{P}_n(\mu_i),
 \]
 \[
\langle { a}_1(\xi_i){\bar
a}_2(\zeta_j)\rangle=\sum_{n=0}^{N_1-1}P_n(\xi_i)S_n(\zeta_j),
 \]
 \[
\langle { a}_2(\mu_i){\bar
a}_1(\eta_j)\rangle=\sum_{n=N_1}^{+\infty}\tilde{S}_n(\eta_j)\tilde{P}_n(\mu_i)={\int\int}
\frac{d\mu(x,y)}{(\eta_j-x)(\mu_i-y)}-\sum_{n=0}^{N_1-1}\tilde{S}_n(\eta_j)\tilde{P}_n(\mu_i),
 \]
 \[
\langle f^{(2)}_{i-1-N-L_2+M_2}{\bar
a}_1(\eta_j)\rangle=\sqrt{h_{N+L_2-M_2-i}}\tilde{S}_{N+L_2-M_2-i}(\eta_j),
 \]
 \[
\langle f^{(2)}_{i-1-N-L_2+M_2}{\bar
a}_2(\zeta_j)\rangle=\sqrt{h_{N+L_2-M_2-i}}S_{N+L_2-M_2-i}(\zeta_j)
 \]

\;

After trivial manipulations with rows and columns of the matrix of
pair-wise v.e.v. we obtain the answer:
 \bea
 {\bf I}_N(\xi, \zeta, \eta, \mu)
 &=&    (-1)^{{1\over 2}(M_1+M_2)(M_1+M_2-1)}(-1)^{L_2 M_1}
 \prod_{n=N}^{{}_{N+L_1-M_1-1}}{\hskip -15 pt}\sqrt{h_n}
 \prod_{n=N}^{{}_{N+L_2- M_2-1}}{\hskip -15 pt}\sqrt{h_n} \cr
 &\& \quad \times
{ \prod_{\alpha=1}^{L_1}\prod_{j=1}^{M_1}(\xi_\alpha -
\eta_j)\prod_{\beta=1}^{L_2} \prod_{k=1}^{M_2}(\zeta_\beta- \mu_k)
 \over \Delta_{L_1}(\xi)  \Delta_{L_2}(\zeta)  \Delta_{M_1}(\eta)  \Delta_{M_2}(\mu)}
 \  \det G,  \cr
 &&
 \label{mainintegral-ref-1a}
 \eea
 where the matrix $G$  is $(L_2+M_1)\times(L_2+M_1)$ matrix which
 consists of six blocks:
$$
\left(\begin{array}{ccccc}
 \ds{\mathop{ K_{11}}^{\!N_1}}(\mu_k, \zeta_\beta)  & \ds{\mathop{ K_{21}}^{\!N_1}}
 (\xi_\alpha, \zeta_\beta)
  & S_{N+L_1-M_1}(\zeta_\beta) & \dots &  S_{N+L_2-M_2-1}(\zeta_\beta) \\
\ds{\mathop{ K_{12}}^{\!N_1}}(\mu_k, \eta_j)
 & \ds{\mathop{ K_{22}}^{\!N_1}}(\xi_\alpha, \eta_j) &
 {\tilde S}_{N+L_1-M_1}(\eta_j) &\dots
   & {\tilde S}_{N+L_2-M_2-1}(\eta_j)\\
\end{array}\right)
$$
where
\[
\alpha=1,\dots,L_1;\quad \beta=1,\dots,L_2;\quad   j= 1,\dots,M_1
;\quad k=1,\dots,M_2
\]
and where  $N_1=N+L_1-M_1$ and
 \be \ds{\mathop{ K_{11}}^{\!J}}(\mu, \zeta) =
\sum_{n=0}^{J-1} {\tilde P}_n(\mu) S_n(\zeta) + {1\over \zeta-
\mu}
\label{wtK22-paper3}\\
 \ee
 \be
 \ds{\mathop{ K_{22}}^{\!J}}(\xi, \eta) =
\sum_{n=0}^{J-1} P_n(\xi) {\tilde S_n}(\eta) + {1\over \xi - \eta}
\label{wtK11-paper3}\\
 \ee
 \be \ds{\mathop{ K_{21}}^{\!J}}(\xi, \zeta) =
\sum_{n=0}^{J-1} P_n(\xi) S_n(\zeta)
\label{K12-paper3}\\
\ee
 \be \ds{\mathop{ K_{12}}^{\!J}}(\mu, \eta) =
\sum_{n=0}^{J-1} {\tilde P}_n(\mu)  {\tilde S}_n(\eta) - \int
{d\mu(x,y) \over (\eta-x)(\mu-y)} \label{wtK21-paper3}
 \ee
 \be\label{01a'-new}
S_{N+L_1-M_1+i-1}(\zeta_j),\quad i=1,\dots,M_1-L_1-M_2+L_2 , \quad
j=1,\dots, L_2
 \ee
 \be\label{02a'-new}
{\tilde S}_{N+L_1-M_1+i-1}(\eta_j) ,\quad
i=1,\dots,M_1-L_1-M_2+L_2,\quad j=1,\dots,M_1,
 \ee

\;

 Now we shall consider six examples, related to the six-block structure, where,
 in each case, only one entry contributes.

\;

 Example 1. $M_1=1$ and $L_1=L_2=M_2=0$, thus $N_2 > N_1 \ge 0$.  We
put $\eta_1=\eta$. In this case the matrix has only one
non-vanishing element, giving the well-known formula
 \be\label{Example-01a}
{\bf I}_N(\eta)=\frac{1}{\sqrt{h_{N-1}}}{\tilde S}_{N-1}(\eta)
 \ee

Example 2. $L_2=1$ and $L_1=M_1=M_2=0$, thus $N_2 > N_1 \ge 0$.
  We put
$\zeta_1=\zeta$. In this case the matrix has only one
non-vanishing element, giving the well-known formula
 \be\label{Example-02a}
{\bf I}_N(\zeta)=\sqrt{h_N}S_{N}(\zeta)
 \ee

\;

Examples below are related to the equality $N_2= N_1\ge 0$.

\;

Example 3. $M_1=L_1=1$ and $L_2=M_2=0$. We put $\xi_1=\xi$ and
$\eta_1=\eta$. In this case the matrix has only one non-vanishing
element and we obtain
 \be\label{Example-11a}
{\bf I}_N(\xi,\eta)= 1+(\xi-{\eta})\sum_{n=0}^{N-1}P_n(\xi){\tilde
S}_n(\eta)
 \ee

Similarly, we have

Example 4.  $L_2=M_2=1$ and $L_1=M_1=0$. In this case the matrix
has only one non-vanishing element, and we obtain
 \be\label{Example-22a}
{\bf I}_N(\zeta,\mu)=
-(\zeta-\mu)\sum_{n=N}^{+\infty}S_n(\zeta){\tilde P}_n(\mu) =1+
(\zeta-\mu)\sum_{n=0}^{N-1}S_n(\zeta){\tilde P}_n(\mu)
 \ee

Example 5. $M_1=M_2=1$ and $L_1=L_2=0$. In this case the matrix
has only one non-vanishing element. We obtain
 \be\label{Example-21a}
{\bf I}_N(\eta,\mu)=\frac{1}{h_{N-1}}\sum_{n=N-1}^{+\infty}{\tilde
S} _n(\eta){\tilde P}_n(\mu)= \frac{1}{h_{N-1}}{\int\int}
\frac{d\mu(x,y)}{(\eta-x)(\mu-y)}-\frac{1}{h_{N-1}}\sum_{n=0}^{N-2}{\tilde
S}_n(\eta){\tilde P} _n(\mu)
 \ee

Example 6.  $L_1=L_2=1$ and $M_1=M_2=0$. In this case the matrix
has only one non-vanishing element and we obtain
 \be\label{Example-12a}
{\bf I}_N(\xi,\zeta)=h_N\sum_{n=0}^{N}P_n(\xi)S_n(\zeta)
 \ee

\;

{\bf Evaluation for the case $N_1\ge N_2\ge 0$ }

This case may be obtained from the previous one, by interchanging
 subscripts $L_1\leftrightarrow L_2$, $M_1\leftrightarrow M_2$,
 $N_1\leftrightarrow N_2$, and $\xi \leftrightarrow
\zeta$,  $\mu \leftrightarrow \eta$  and also  $P_n
\leftrightarrow S_n$, ${\tilde P}_n \leftrightarrow {\tilde S}_n$.

We obtain the answer which coincides with the answer given by
formulae (1.8)-(1.16) of \cite{paper2}:
 \bea
 {\bf I}_N(\xi, \zeta, \eta, \mu)
 &=& (-1)^{{1\over 2}(M_1+M_2)(M_1+M_2-1)}(-1)^{L_1 M_2}
 \prod_{n=0}^{N-1}h_n^{-1}\prod_{n=N}^{{}_{N+L_2-M_2-1}}{\hskip -15 pt}\sqrt{h_n}
 \prod_{n=N}^{{}_{N+L_1- M_1-1}}{\hskip -15 pt}\sqrt{h_n} \cr
 &\& \quad \times
{ \prod_{\alpha=1}^{L_1}\prod_{j=1}^{M_1}(\xi_\alpha -
\eta_j)\prod_{\beta=1}^{L_2} \prod_{k=1}^{M_2}(\zeta_\beta- \mu_k)
 \over \Delta_{L_1}(\xi)  \Delta_{L_2}(\zeta)  \Delta_{M_1}(\eta)  \Delta_{M_2}(\mu)}
 \  \det G,  \cr
 &&
 \label{mainintegral-paper2}
 \eea
where $(L_1+M_2)\times(L_1+M_2)$ matrix $G$ consists of six
blocks:
$$
\left(\begin{array}{ccccc}
 \ds{\mathop{ K_{11}}^{\!N_2}}(\xi_\alpha, \eta_j)
 & \ds{\mathop{ K_{12}}^{\!N_2}}(\xi_\alpha, \zeta_\beta)
  & P_{N+L_2-M_2}(\xi_\alpha)
  & \dots
  &  P_{N+L_1-M_1-1}(\xi_\alpha) \\
  \ds{\mathop{ K_{21}}^{\!N_2}}(\mu_k, \eta_j)
  & \ds{\mathop{ K_{22}}^{\!N_2}}(\mu_k, \zeta_\beta)
  &   {\tilde P}_{N+L_2-M_2}(\mu_k)&\dots
   & {\tilde P}_{N+L_1-M_1-1}(\mu_k)\\
\end{array}\right)
$$
where
\[
\alpha=1,\dots,L_1;\quad \beta=1,\dots,L_2;;\quad   j= 1,\dots,M_1
;\quad k=1,\dots,M_2
\]
and where  $N_2=N+L_2-M_2$ and
 \be
 \ds{\mathop{ K_{11}}^{\!J}}(\xi, \eta) =
\sum_{n=0}^{J-1} P_n(\xi) {\tilde S_n}(\eta) + {1\over \xi - \eta}
\label{wtK11-paper2}\\
 \ee
 \be \ds{\mathop{ K_{22}}^{\!J}}(\mu, \zeta) =
\sum_{n=0}^{J-1} {\tilde P}_n(\mu) S_n(\zeta) + {1\over \zeta-
\mu}
\label{wtK22-paper2}\\
 \ee
 \be \ds{\mathop{ K_{12}}^{\!J}}(\xi, \zeta) =
\sum_{n=0}^{J-1} P_n(\xi) S_n(\zeta)
\label{K12-paper2}\\
\ee
 \be \ds{\mathop{ K_{21}}^{\!J}}(\mu, \eta) =
\sum_{n=0}^{J-1} {\tilde P}_n(\mu)  {\tilde S}_n(\eta) - \int
{d\mu(x,y) \over (\eta-x)(\mu-y)} \label{wtK21-paper2}
 \ee

Example 7. $M_2=1$ and $L_2=L_1=M_1=0$, thus $N_1 > N_2 \ge 0$. We
put $\mu_1=\mu$. Then
 \be\label{Example-02a'}
{\bf I}_N(\mu)=\frac{1}{\sqrt{h_{N-1}}}{\tilde P}_{N-1}(\mu)
 \ee

Example 8. $L_1=1$ and $L_2=M_2=M_1=0$, thus $N_1 > N_2 \ge 0$.
 We put $\xi_1=\xi$. We obtain the well-known formula
 \be\label{Example-01a'}
{\bf I}_N(\xi)=\sqrt{h_{N}}P_{N}(\xi)
 \ee

\subsection{When $N_1<0$}

In cases listed in the Introduction as (2) and (3) we have
$N_1<0$. Let us explicitly write down
\begin{equation}\label{vac-case2}
N_1\le 0\le N_2:\quad \langle N_1,-N_2|=\langle 0,0| {
f}^{(1)}_{-1}\cdots { f}^{(1)}_{N_1}{ f}^{(2)}_{-1}\cdots{
f}^{(2)}_{-N_2} \ ,
\end{equation}
\begin{equation}\label{vac-case3}
N_1\le N_2\le 0:\quad \langle N_1,-N_2|=\langle 0,0| {
f}^{(1)}_{-1}\cdots { f}^{(1)}_{N_1} {\bar f}^{(2)}_{0}\cdots
{\bar f}^{(2)}_{-N_2-1}
\end{equation}
Then, as it follows from (\ref{g-e-f-1})-(\ref{g-e-f-2}), in
either case
\begin{equation}\label{langle-Q}
\langle N_1,-N_2| Q^{-1}=\langle N_1,-N_2|
\end{equation}
Thus, in the both cases, we re-write (\ref{mainequality-B}) as
\begin{equation}\label{step3-b-}
 {\bf I}_N(\xi, \zeta, \eta, \mu)R_N^{-1}=
\langle N_1,-N_2|{B}_2(\mu)B_1(\xi){\bar B}_1(\eta){\bar
B}_2(\zeta) |0,0\rangle
\end{equation}

\subsection{Evaluation for the case (2): $N_1\le 0\le N_2$.}

Thus, by (\ref{step3-b-}) and (\ref{vac-case2}) we have
\begin{equation}\label{step3-b}
 {\bf I}_N(\xi, \zeta, \eta, \mu)R_N^{-1} =
\langle 0,0| { f}^{(1)}_{-1}\cdots { f}^{(1)}_{N_1}{
f}^{(2)}_{-1}\cdots{ f}^{(2)}_{-N_2} {B}_2(\mu)B_1(\xi){\bar
B}_1(\eta){\bar B}_2(\zeta) |0,0\rangle
\end{equation}

By Wick's formula (\ref{Wick-det}) the right hand side is the
determinant of a $L_2+M_1$ by $L_2+M_1$ matrix which consists of
eight blocks:
$$
\left(\begin{array}{cccc}
 \langle { b}_1(\xi_\alpha){\bar b}_1(\eta_j)\rangle
 & \langle{ b}_2(\mu_k){\bar b}_1 (\eta_j)\rangle
 & \langle f^{(2)}_{b-1-N-L_2+M_2}{\bar b}_1(\eta_j)\rangle
 & \langle f^{(1)}_{m-1+N+L_1-M_1}{\bar b}_1(\eta_j)\rangle\\
  \langle{ b}_1(\xi_\alpha){\bar b}_2 (\zeta_\beta)\rangle
   & \langle{ b}_2(\mu_k){\bar b}_2 (\zeta_{\beta})\rangle
   & \langle f^{(2)}_{b-1-N-L_2+M_2}{\bar b}_2(\zeta_\beta)\rangle
   & \langle f^{(1)}_{m-1+N+L_1-M_1}{\bar b}_2(\zeta_\beta)\rangle\\
\end{array}\right)
$$
where
\[
\alpha=1,\dots,L_1;\quad \beta=1,\dots,L_2;\quad   j= 1,\dots,M_1
;\quad k=1,\dots,M_2;
\]
\[
b=1,\dots,N+L_2-M_2;\quad m=1,\dots,-N-L_1+M_1
\]

Then using (\ref{b1})-(\ref{bar-b2}), we evaluate all pair-wise
vacuum expectation values:
\[
\langle { b}_1(\xi_\alpha){\bar b}_1(\eta_j)\rangle=
\frac{1}{\xi_\alpha}\frac{1}{1-\frac{\eta_j}{\xi_\alpha}}=\frac{1}{\xi_\alpha-\eta_j},
\]
\[
\langle{ b}_2(\mu_k){\bar b}_2
(\zeta_\beta)\rangle=\sum_{n=0}^{+\infty}\tilde{P}_n(\mu_k)S_n(\zeta_\beta)
=\frac{1}{\mu_k-\zeta_\beta} ,
\]
\[
\langle{ b}_2(\mu_k){\bar b}_1 (\eta_j)\rangle=
\sum_{n=0}^{+\infty}\tilde{P}_n(\mu_k)\tilde{S}_n(\eta_j)=H(\mu_k,\eta_j)
\]
\[
\langle{ b}_1(\xi_\alpha){\bar b}_2 (\zeta_\beta)\rangle= 0,
\]
\[
\langle f^{(2)}_{b-1-N-L_2+M_2}{\bar
b}_1(\eta_j)\rangle=\sqrt{h_{N+L_2-M_2-b}}\tilde{S}_{N+L_2-M_2-b}(\eta_j)
,\quad b=1,\dots,N+L_2-M_2,
\]
\[
\langle f^{(2)}_{b-1-N-L_2+M_2}{\bar
b}_2(\zeta_\beta)\rangle=\sqrt{h_{N+L_2-M_2-b}}S_{N+L_2-M_2-b}(\zeta_\beta)
,\quad b=1,\dots,N+L_2-M_2,
\]
\[
\langle f^{(1)}_{m-1+N+L_1-M_1}{\bar
b}_1(\eta_j)\rangle=\eta^{-N-L_1+M_1-m}_j ,\quad
m=1,\dots,-N-L_1+M_1,
\]
\[
\langle f^{(1)}_{m-1+N+L_1-M_1}{\bar b}_2(\zeta_\beta)\rangle=0,
\quad m=1,\dots,-N-L_1+M_1
\]

 After trivial manipulations with rows and columns of the
matrix of pair-wise v.e.v. we obtain the answer
\[
{\bf I}_N (\xi, \zeta, \eta, \mu) =
\]
 \be\label{answer-b-paper2}
=\epsilon\prod_{n=N}^{{}_{N+L_1-M_1-1}}{\hskip -15 pt}\sqrt{h_n}
 \prod_{n=N}^{{}_{N+L_2- M_2-1}}{\hskip -15 pt}\sqrt{h_n}
\frac{ \prod_{\alpha=1}^{L_1}\prod_{j=1}^{M_1}(\xi_\alpha -
\eta_j)\prod_{\beta=1}^{L_2}\prod_{k=1}^{M_2}(\zeta_\beta- \mu_k)
 }
 {\Delta_{L_1}(\xi)\Delta_{L_2}(\zeta)\Delta_{M_1}(\eta)\Delta_{M_2}(\mu)}
\det G ,
 \ee
\[
\epsilon= (-1)^{{1\over 2}N(N-1)+N(L_1-M_1)+{1\over 2}M_1(M_1-1)+
{1\over 2}L_2(L_2-1)+L_1(L_2-M_2)+L_2M_2-M_2}
\]
where the  $(L_2+M_1)\times(L_2+M_1)$ matrix $G$ consists of eight
blocks
 \be\label{case2}
\left(\begin{array}{cccc}
  {1\over \mu_k-  \zeta_\beta} & 0 & S_b(\zeta_\beta) & 0\\
  H(\mu_k,\eta_j) & {1\over \xi_\alpha-\eta_j} & {\tilde S}_b(\eta_j)
   & { S}_{m}(\eta_j)\\
\end{array}\right)
 \ee
where
\[
\alpha=1,\dots,L_1;\quad \beta=1,\dots,L_2;\quad   j= 1,\dots,M_1
;\quad k=1,\dots,M_2;
\]
\[
b=1,\dots,N+L_2-M_2;\quad m=1,\dots,-N-L_1+M_1
\]

\;

{\bf Evaluation for the case  $N_2\le 0\le N_1$.}

The answer for this case may be obtained by the answer for the
previous case, if we inter-change
 subscripts $L_1\leftrightarrow L_2$, $M_1\leftrightarrow M_2$,
 $N_1\leftrightarrow N_2$, and the variables $\xi \leftrightarrow
\zeta$,  $\mu \leftrightarrow \eta$ and also  $P_n \leftrightarrow
S_n$, ${\tilde P}_n \leftrightarrow {\tilde S}_n$, and (due to the
definition (\ref{sum-s*p*})) $H(\mu,\eta)\to H(\mu,\eta)$ .

The answer coincides with the answer of \cite{paper2} (see
(3.21)-(3.22) there):
\[
{\bf I}_N(\xi, \zeta, \eta, \mu) =
\]
 \be\label{answer-b-paper2''}
=\epsilon\prod_{n=N}^{{}_{N+L_2-M_2-1}}{\hskip -15 pt}\sqrt{h_n}
 \prod_{n=N}^{{}_{N+L_1- M_1-1}}{\hskip -15 pt}\sqrt{h_n}
\frac{ \prod_{\alpha=1}^{L_1}\prod_{j=1}^{M_1}(\xi_\alpha -
\eta_j)\prod_{\beta=1}^{L_2}\prod_{k=1}^{M_2}(\zeta_\beta- \mu_k)
 }
 {\Delta_{L_1}(\xi)\Delta_{L_2}(\zeta)\Delta_{M_1}(\eta)\Delta_{M_2}(\mu)}
\det G ,
 \ee
\[
\epsilon= (-1)^{{1\over 2}N(N-1)+N(L_2-M_2)+{1\over 2}M_2(M_2-1)+
{1\over 2}L_1(L_1-1)+L_2(L_1-M_1)+L_1M_1-M_1}
\]
where the $(L_1+M_2)\times(L_1+M_2)$ matrix $G$ consists of eight
blocks
 \be\label{case2''}
\left(\begin{array}{cccc}
   {1\over \xi_\alpha-\eta_i}& 0 & P_b(\xi_\alpha) & 0\\
  H(\mu_k,\eta_i) & {1\over \mu_k - \zeta_\beta } & {\tilde P}_{b}(\mu_k)
   & { S}_{m}(\mu_k)\\
\end{array}\right)
 \ee
where
\[
\alpha=1,\dots,L_1;\quad   \beta= 1,\dots,L_2;\quad i=1,\dots,M_1
; \quad k=1,\dots,M_2;
\]
\[
b=1,\dots,N+L_1-M_1 ;\quad m=1,\dots, M_2-L_2-N
\]

\subsection{Evaluation for the case (3): $N_1\le N_2\le 0$ }

In this case we have (\ref{step3-b-}).

Using transformation (\ref{shift1})-(\ref{shift2}) which leaves
vacuum expectation values invariant,  in form:
\[
\langle *,*|\to\langle  *,*+N_2|,\quad |*,*\rangle \to
|*,*+N_2\rangle ,
\]
\[
f^{(2)}_{i}\to f^{(2)}_{i+N_2},\quad {\bar f}^{(2)}_{i}\to {\bar
f}^{(2)}_{i+N_2},\quad i\in\mathbb{Z},
\]
we write
\[
 {\bf I}_N(\xi, \zeta, \eta, \mu)R_N^{-1}=
\langle N_1,-N_2|{B}_2(\mu)B_1(\xi){\bar B}_1(\eta){\bar
B}_2(\zeta) |0,0\rangle
\]
\begin{equation}\label{step2'''}
=\langle N_1,0| { C}_2(\mu)C_1(\xi){\bar C}_1(\eta){\bar
C}_2(\zeta) |0,N_2\rangle
\end{equation}
where we shift the charges of the second component of the vacuum
vector by $N_2$ and replaced fermionic operators $b_1(\xi),{\bar
b}_1(\eta),b_2(\mu)$ and ${\bar b}_2(\zeta)$ given by
(\ref{b1})-(\ref{bar-b2}), by
 \be\label{c1}
c_1(\xi)=
\sum_{n=-\infty}^{+\infty}{f}_{n}^{(1)}P_n(\xi)\sqrt{h_n} \ ,
 \ee
 \be\label{bar-c1}
c_1(\eta)=  \sum_{n=-\infty}^{+\infty}{\bar
f}_n^{(1)}{\tilde{S}_n(\eta)\over\sqrt{h_n}}+\sum_{n=0}^{+\infty}{\bar
f}_{-n-1+N_2}^{(2)}{\tilde{S}_n(\eta)\sqrt{h_n}} \ ,
 \ee
 \be\label{c2}
c_2(\mu)=
\sum_{n=-\infty}^{+\infty}f_{-n-1+N_2}^{(2)}{\tilde{P}_n(\mu)\over\sqrt{h_n}}
-\sum_{n=0}^{+\infty}f_{n}^{(1)}{\tilde{P}_n(\mu)\sqrt{h_n}} \ ,
 \ee
 \be\label{bar-c2}
{\bar c}_2(\zeta)= \sum_{n=-\infty}^{+\infty}{\bar
f}_{-n-1+N_2}^{(2)}S_n(\zeta)\sqrt{h_n}
 \ee

We produced this shift of the vacuum charge and the fermion
numbering in order to come to the form of v.e.v., where all
fermions with bar are situated to the right:
\begin{equation}\label{step2''''}
\langle 0,0|{ f}^{(1)}_{-1}\cdots { f}^{(1)}_{N_1}
\prod_{n=1}^{M_2}{
c}_2(\mu_n)\prod_{n=1}^{L_1}c_1(\xi_n)\prod_{n=1}^{M_1}{\bar
c}_1(\eta_n)\prod_{n=1}^{L_2}{\bar c}_2(\zeta_n) {\bar
f}^{(2)}_{N_2}\cdots
  {\bar f}^{(2)}_{-1}|0,0\rangle ,
\end{equation}
where  formula (\ref{Wick-det}) is applicable. We have nine-block
matrix:
$$
\left(\begin{array}{ccc}
 \langle {c}_1(\xi_\alpha){\bar c}_1(\eta_j)\rangle
 & \langle{ c}_1(\xi_\alpha){\bar c}_2 (\zeta_\beta)\rangle
 & \langle {c}_1(\xi_\alpha){\bar f}^{(2)}_{m-1+N+L_2-M_2}\rangle\\
  \langle{ c}_2(\mu_k){\bar c}_1 (\eta_j)\rangle
  & \langle{ c}_2(\mu_k){\bar c}_2(\zeta_\beta)\rangle
  & \langle { c}_2(\mu_k){\bar f}^{(2)}_{m-1+N+L_2-M_2}\rangle \\
   \langle f^{(1)}_{\ell-1+N+L_1-M_1}{\bar c}_1(\eta_j)\rangle
   & \langle f^{(1)}_{\ell-1+N+L_1-M_1}{\bar c}_2(\zeta_\beta)\rangle
   &0\\
\end{array}\right)
$$
where
\[
\alpha=1,\dots,L_1;\quad   \beta= 1,\dots,L_2;\quad j=1,\dots,M_1
; \quad k=1,\dots,M_2;
\]
\[
\ell=1,\dots,-N-L_1+M_1;\quad m=1,\dots,N+L_2-M_2
\]

Pair-wise expectation values are
\[
\langle { c}_1(\xi){\bar c}_1(\eta)\rangle=
\frac{1}{\xi}\frac{1}{1-\frac{\eta}{\xi}}=\frac{1}{\xi-\eta},
\]
\[
\langle{ c}_2(\mu){\bar c}_2
(\zeta)\rangle=\sum_{n=0}^{+\infty}\tilde{P}_{n}(\mu)S_{n}(\zeta)
=\frac{1}{\mu-\zeta},
\]
\[ \langle{ c}_2(\mu){\bar c}_1 (\eta)\rangle=
\sum_{n=0}^{+\infty}\tilde{P}_{n}(\mu)\tilde{S}_{n}(\eta)=H(\mu,\eta)
,
\]
\[
\langle{ c}_1(\xi){\bar c}_2 (\zeta)\rangle= 0,
\]
\[
\langle f^{(1)}_{\ell-1+N+L_1-M_1}{\bar
 c}_1(\eta)\rangle=\tilde{S}_{\ell-1+N+L_1-M_1}(\eta)\sqrt{h_{\ell-1+N+L_1-M_1}}=
\eta^{-N-L_1+M_1-\ell},\quad \ell=1,\dots,-N-L_1+M_1,
\]
\[
\langle f^{(1)}_{\ell-1+N+L_1-M_1}{\bar c}_2(\zeta)\rangle=0,
\quad \ell=1,\dots,-N-L_1+M_1 ,
\]
\[
\langle {c}_1(\xi){\bar f}^{(2)}_{m-1+N+L_2-M_2}\rangle=0 ,\quad
m=1,\dots,N+L_2-M_2,
\]
\[
\langle { c}_2(\mu){\bar
f}^{(2)}_{m-1+N+L_2-M_2}\rangle=\tilde{P}_{-m}(\mu)\sqrt{h_{-m}}=\mu^{\ell-1}
,\quad m=1,\dots,N+L_2-M_2
\]

 After trivial manipulations with rows and columns of the
matrix of pair-wise v.e.v. we obtain the answer which coincides
with the answer of \cite{paper2} (given by (3.35)-(3.36) there):
\[
{\bf I}_N (\xi, \zeta, \eta, \mu) =
\]
 \be\label{answer-c-paper2*}
=\epsilon\prod_{n=N}^{{}_{N+L_2-M_2-1}}{\hskip -15 pt}\sqrt{h_n}
 \prod_{n=N}^{{}_{N+L_1- M_1-1}}{\hskip -15 pt}\sqrt{h_n}
\frac{ \prod_{\alpha=1}^{L_1}\prod_{j=1}^{M_1}(\xi_\alpha -
\eta_j)\prod_{\beta=1}^{L_2}\prod_{k=1}^{M_2}(\zeta_\beta- \mu_k)
 }
 {\Delta_{L_1}(\xi)\Delta_{L_2}(\zeta)\Delta_{M_1}(\eta)\Delta_{M_2}(\mu)}
\det G ,
 \ee
\[
\epsilon=(-1)^{\frac12 L_2(L_2-1)+\frac12 M_1(M_1-1)+\frac12
N(N+1)+L_2(L_1+M_1)+N(L_1+M_1)+M_2L_2}
\]
where    $G$ is $(M_1+M_2-N)\times(M_1+M_2-N) $ matrix which
consists of nine blocks:
 \be\label{case3}
\left(\begin{array}{ccc}
   \frac{1}{\xi_\alpha-\eta_j} & 0 & 0\\
  H(\mu_k,\eta_j) & \frac{1}{\mu_k-\zeta_\beta} & S_{m}(\mu_k)
   \\ P_{\ell}(\eta_j)&0 &0\\
\end{array}\right)
 \ee
where
\[
\alpha=1,\dots,L_1;\quad \beta=1,\dots,L_2;\quad
j=1,\dots,M_1;\quad k=1,\dots,M_2;
\]
\[
\ell=1,\dots,M_1-L_1-N;\quad m=1,\dots,N+L_2-M_2
\]

\;

{\bf Evaluation for the case $N_2\le N_1\le 0$}

After replacements we obtain
\[
{\bf I}_N(\xi, \zeta, \eta, \mu) =
\]
 \be\label{answer-c-paper2''}
=\epsilon\prod_{n=N}^{{}_{N+L_2-M_2-1}}{\hskip -15 pt}\sqrt{h_n}
 \prod_{n=N}^{{}_{N+L_1- M_1-1}}{\hskip -15 pt}\sqrt{h_n}
\frac{ \prod_{\alpha=1}^{L_1}\prod_{j=1}^{M_1}(\xi_\alpha -
\eta_j)\prod_{\beta=1}^{L_2}\prod_{k=1}^{M_2}(\zeta_\beta- \mu_k)
 }
 {\Delta_{L_1}(\xi)\Delta_{L_2}(\zeta)\Delta_{M_1}(\eta)\Delta_{M_2}(\mu)}
\det G ,
 \ee
\[
\epsilon=(-1)^{\frac12 L_1(L_1-1)+\frac12 M_2(M_2-1)+\frac12
N(N+1)+L_1(L_2+M_2)+N(L_2+M_2)+M_1L_1}
\]
 where $G$ is $(M_1+M_2-N)\times(M_1+M_2-N) $ matrix which consists of nine blocks
 \be\label{case3''}
\left(\begin{array}{ccc}
   \frac{1}{\zeta_\beta-\mu_k} & 0 & 0\\
  H(\mu_k,\eta_j) & \frac{1}{\eta_j-\xi_\alpha} & P_m(\eta_j)
   \\ S_{\ell}(\mu_k)&0 &0\\
\end{array}\right)
 \ee
where
\[
\alpha=1,\dots,L_1;\quad \beta=1,\dots,L_2;\quad j=1,\dots,M_1;
\quad k=1,\dots,M_2;
\]
\[
\ell=1,\dots,M_2-L_2-N  ;\quad m=1,\dots, M_1-L_1-N
\]

\section*{Acknowledgements}

The authors  would like to thank T. Shiota and J. van de Leur for
helpful discussions, and  (A.O.)  thanks A. Odzijevicz for kind
hospitality during his stay in Bialystok in June 2005.

\appendix

\section{Appendix A}

By (\ref{2-fermions-antic}) we have for $ p,q \in\mathbb{C}$:
 \be\label{g-e-1}
[(g_n)^{p},(g_m)^{q}]=0,\quad [(e_n)^{p},(e_m)^{q}]=0,\quad
n,m\in\mathbb{Z}
 \ee
and
 \be\label{g-e-2}
[(g_n)^{p},(e_m)^{q}]=0,\quad n\neq m
 \ee
where $[,]$ denotes the commutator.

Also:
 \be\label{g-e-f-1}
[(g_n)^{p},f_m^{(1)}] =[(g_n)^{p},{\bar f}_m^{(2)}]=
[(e_n)^{p},{\bar f}_m^{(1)}]=[(e_n)^{p},{ f}_m^{(2)}]=0,\quad
n,m\in\mathbb{Z},
 \ee
 \be\label{g-e-f-2}
[(g_n)^{p},{\bar f}_m^{(1)}] =[(g_n)^{p},{ f}_{-m-1}^{(2)}]=
[(e_n)^{p},{ f}_m^{(1)}]=[(e_n)^{p},{\bar f}_{-m-1}^{(2)}]=0
,\quad n\neq m,
 \ee
and
 \be\label{g-f-3'}
(g_n)^{-p}{\bar f}_n^{(1)}(g_n)^{p}={\bar f}_n^{(1)} +p {\bar
f}_{-n-1}^{(2)}, \quad (g_n)^{-p}{ f}_{-n-1}^{(2)}(g_n)^{p}={
f}_{-n-1}^{(2)}-p { f}_{n}^{(1)} ,\quad n\in\mathbb{Z},
 \ee
 \be\label{e-f-3'}
(e_n)^{p}{ f}_n^{(1)}(e_n)^{-p}={ f}_n^{(1)}-p { f}_{-n-1}^{(2)}
,\quad (e_n)^{p}{\bar f}_{-n-1}^{(2)}(e_n)^{-p}={\bar
f}_{-n-1}^{(2)}+p {\bar f}_{n}^{(1)}\quad n\in\mathbb{Z}
 \ee

\section{Appendix B}

For $ {\hat \omega}:=\sum_{n,m} \omega_{n,m}f_n{\bar f}_m $ we
have
\[
{\textsl{ad} }_{\hat \omega} \; f_i=\sum_n f_n\omega_{n,i}\;
,\qquad {\textsl{ad} }_{\hat \omega} \; {\bar f}_i=-\sum_n
\omega_{i,n}{\bar f}_n
\]
Then it follows
\[
{\textsl{Ad} }_{\hat \omega} \; f_i=\sum_n f_n\left(e^\omega
\right)_{n,i}\; ,\qquad {\textsl{Ad} }_{\hat \omega} \;  {\bar
f}_i=\sum_n \left(e^{-\omega} \right)_{i,n}{\bar f}_n
\]
This yields (\ref{d1}) and (\ref{d1bar}) which are equivalent
respectively to (\ref{d-f-p}) and (\ref{d-f-p*}).

In the same way we prove (\ref{d2bar}) and (\ref{d2}) which are
equivalent respectively to (\ref{d-f-s}) and (\ref{d-f-s*}).

\section{Appendix C}

Now, let us show that
 \be\label{p*-s-s*-p'}
{\tilde S}_n(\eta)={\int\int} \frac{S_n(y)}{\eta-x}d\mu(x,y)
,\quad {\tilde P}_n(\mu) ={\int\int}
\frac{P_n(x)}{\mu-y}d\mu(x,y),\quad n\ge 0
 \ee

\;

Proof I.
 \be\label{sum-pp*}
[ \sum_{n=0}^{+\infty}d^{(1)}_n x^n,\sum_{n=0}^{+\infty}{\bar
d}^{(1)}_n \eta^{-n-1}]_+=\frac{1}{\eta}\frac{1}{1-\frac x\eta}=
\sum_{n=0}^{+\infty} P_n(x)\tilde{S}_n(\eta)
 \ee
where the first equality follows from the definitions (\ref{d1})
and (\ref{d1bar}), while the equality of the anti-commutator to
the last member follows from (\ref{d-f-p})-(\ref{d-f-p*}).
Multiplying both sides of the second equality by $S_n(y)d\mu(x,y)$
and integrating we come to the first equality (\ref{p*-s-s*-p'}).

We obtain the second equality (\ref{p*-s-s*-p'}) by the similar
integration of
 \be\label{sum-ss*}
 [ \sum_{n=0}^{+\infty}d^{(2)}_{-n-1}
\mu^{-n-1},\sum_{n=0}^{+\infty}{\bar d}^{(2)}_{-n-1}
y^{n}]_+=\frac{1}{\mu}\frac{1}{1-\frac y\mu}= \sum_{n=0}^{+\infty}
S_n(y)\tilde{P}_n(\mu)
 \ee
(These equalities results from definitions (\ref{d2}) and
(\ref{d2bar}), and  from (\ref{d-f-s*}),(\ref{d-f-s})).

 We shall also use (in
(\ref{Example-21a}) below) the corollary of these equalities:
 \be\label{sum-s*p*}
{\int\int}
\frac{d\mu(x,y)}{(\eta-x)(\mu-y)}=\sum_{n=0}^{\infty}{\tilde
S}_n(\eta){\tilde P} _n(\mu)=:H(\mu,\eta)
 \ee
 which is obtained by multiplying of the right hand sides of the
 relations (\ref{sum-pp*}) and
 (\ref{sum-ss*}), integrating and using the orthogonality
 of $P_n$ and $S_n$.

\;

Proof II. Considering the sum of entries weighted with powers of
$x^{-m-1}$ (where $m$ is the row-number of the entry) of the n-th
column of relation $K^{-1}H=B{\bar K}^{-1}$ (which follows from
the known factorization relation $H{\bar K}=KB$), we obtain the
first equality of (\ref{p*-s-s*-p'}) from the first of
(\ref{K-p*-K-s*}) and from the definition of bi-moments. (The
second relation in  (\ref{p*-s-s*-p'}) is proved similarly).

\section{Appendix: Links with soliton theory}
\renewcommand{\theequation}{\Alph{section}-\arabic{equation}}
\setcounter{equation}{0}

In this appendix, we discuss the links between the evaluation of
such symmetric rational integrals and soliton theory. There are
two problems:

(A) To find which measure $d\mu$ is related to a given matrix
integral? We do not know the general answer to this problem. A
partial answer  can be found in the papers \cite{ZJZ}, \cite{1'}
and  the Appendices to \cite{OS} and to \cite{paper1}.

(B) To find which measure  $d\mu$ is related to soliton theory.
This problem is addressed in \cite{paper1} and \cite{paper5}.

We restrict ourselves to the relation to the usual (one-component)
TL hierarchy.

Indeed, one can easily show that the expression
 \be\label{2-comp-ferm-TL-tau}
\tau_N({\bf t}^{(1)},{\bf t}^{(2)})=\langle N,-N|  g({\bf
t}^{(1)},{\bf t}^{(2)})|0,0\rangle ,
 \ee
where $g({\bf t}^{(1)},{\bf t}^{(2)})=e^{A({\bf t}^{(1)},{\bf
t}^{(2)})}$, and $A $ is of form
 \be
A({\bf t}^{(1)},{\bf t}^{(2)})={\int\int}\sum_{i,j=1,2}
e^{V(x,{\bf t}^{(1)})-V(y,{\bf t}^{(2)})}f^{(1)}(x){\bar
f}^{(2)}(y)d\mu_{0}(x,y)
 \ee
where
 \be
V(x,{\bf t}^{(1)})=\sum_{n=0}^\infty t_n^{(1)}x^n ,\quad V(y,{\bf
t}^{(2)})=\sum_{n=0}^\infty t_n^{(2)}y^n
 \ee
and where $d\mu_{0}(x,y)$ is a rather arbitrary measure, fits into
a general expression for  tau functions of the two-component KP
hierarchy and also the TL hierarchy developed in \cite{DJKM} (see
also \cite{UT}, \cite{KdL}).

 Remark: the notations $ t_m^{(1)}$ and $ t_m^{(2)}$ are  related
respectively to the notations $-\frac{u_m}{m}$ and $\frac{v_m}{m}$
of \cite{paper1}. Our $V(x,{\bf t}^{(j)})$ here is $V_j(x)$ of
\cite{paper1}.

In \cite{paper1} it is shown that the expression
(\ref{2-comp-ferm-TL-tau}) on the one hand gives rise to multiple
integrals, which, in turn, for certain choices of the measure
$d\mu_{0}(x,y)$ (see Problem (A) above)  can be identified with a
partition function of some model of random matrices. On the other
hand it is also an example of TL tau function.

This means that if we choose
 \be\label{measure-deformed}
d\mu(x,y)=d\mu(x,y,{\bf t}^{(1)},{\bf t}^{(2)})=e^{V(x,{\bf
t}^{(1)})-V(y,{\bf t}^{(2)})}d\mu_{0}(x,y)
 \ee
then the expectation value (\ref{2-comp-ferm-TL-tau}) is a TL (and
also two-component KP) tau function.

 Remark: if $d\mu_0$ solves this or that differential equation,
than, one can write a correspondent "string equation".

Notice that the first non-trivial member of the TL hierarchy was
introduced and integrated in \cite{AM} and called the relativistic
two-dimensional Toda lattice. In this paper a factorization
problem similar to (\ref{factorization}) (but with matrices that
are infinite in both directions) was considered (see also
\cite{UT}).

Notational Remark. In formulas below we shall denote the pair of
sets of  times $({\bf t}^{(1)},{\bf t}^{(2)})$ by a single letter
${\bf t}$. The only quantity which depends on a single such set
(either ${\bf t}^{(1)}$ or ${\bf t}^{(2)}$) is $V$, where we shall
point out which set it depends on.

Then from general consideration in \cite{DJKM} (see also
\cite{UT}, \cite{KdL}) one finds that the integrals of rational
functions $I_N(\eta,\xi, \mu,\zeta)$, see
(\ref{M-1-L1-M-2-L2-insertions}), may be obtained via the
bosonization procedure as follows
\[
{\bf I}_N (\xi, \zeta, \eta, \mu ;{\bf t})=
\]
 \be\label{bosonization}
\tau_N({\bf t})^{-1}\prod_{i=1}^{L_1}\xi_i^ND_1(\xi_i)
\prod_{i=1}^{M_1}\eta_i^{-N}D_1(\eta_i)^{-1}
\prod_{i=1}^{M_2}\mu_i^{-N}D_2(\mu_i)
\prod_{i=1}^{L_2}\zeta_i^ND_2(\zeta_i)^{-1}\tau_N({\bf t}),
 \ee
where
 \be\label{shift}
D_j(z)=\exp \ \left(- \sum_{n=1}^\infty \frac {1}{nz^{n}}
\frac{\partial}{\partial t_n^{(j)}}\right)
 \ee
is a vertex operator.

Notational Remark. In \cite{paper2} we use the notations
 \be\label{Pp'}
P_n(x)=\frac{1}{\sqrt{h_n}}p_n(x),\quad
 \ee
 \be\label{P*p*'}
{\tilde S}_{n}(x)={\sqrt{h_n}}p_{n}^*(x)
 \ee
 \be\label{S*s*'}
{\tilde P}(y)={\sqrt{h_n}}s_{n}^*(y,{\bf t}),
 \ee
 \be\label{Ss'}
S_n(y)=\frac{1}{\sqrt{h_n}}s_n(y)
 \ee

The functions
 \be\label{Pp}
\psi_n^{(1)}(x,{\bf t}):=e^{V(x,{\bf t}^{(1)})}P_n(x,{\bf
t})\sqrt{h_n}=e^{V(x,{\bf t}^{(1)})}p_n(x,{\bf t}),\quad
 \ee
 \be\label{P*p*}
\psi_n^{*(1)}(x,{\bf t}):=e^{-V(x,{\bf t}^{(1)})}\frac{{\tilde
S}_{n-1}(x,{\bf t})}{\sqrt{h_{n-1}}}=e^{-V(x,{\bf
t}^{(1)})}p_{n-1}^*(x,{\bf t})
 \ee
are to be interpreted as first component Baker functions
(respectively, adjoint first component Baker functions) which
depend on a spectral parameter $x$, while
 \be\label{S*s*}
\psi_n^{(2)}(y,{\bf t}):=e^{V(y,{\bf t}^{(2)})}\frac{{\tilde
P}_{n-1}(y,{\bf t})}{\sqrt{h_{n-1}}}=e^{V(y,{\bf
t}^{(2)})}s_{n-1}^*(y,{\bf t}),
 \ee
 \be\label{Ss}
\psi_n^{*(2)}(y,{\bf t}):= e^{-V(y,{\bf t}^{(2)})}S_n(y,{\bf
t})\sqrt{h_n}=e^{-V(y,{\bf t}^{(2)})}s_n(y,{\bf t})
 \ee
are to be interpreted as second component Baker functions
(respectively, adjoint second component Baker functions) which
depend on a spectral parameter $y$.

This fact is due to the formulae (notice the factor $(-1)^N$ which
result from  re-ordering the fermions of two different types to
achieve true sign according to (\ref{2-vacuum}))
 \be\label{P-via-vertex-tau}
\psi_n^{(1)}(x,{\bf t})=(-1)^Ne^{V(x,{\bf t}^{(1)})}\frac{\langle
N+1,-N|f^{(1)}(x) g({\bf t})|0,0\rangle}{\langle N,-N| g({\bf
t})|0,0\rangle}= x^{N}e^{V(x,{\bf
t}^{(1)})}\frac{D_1(x)\tau_N({\bf t})}{\tau_N({\bf t})}
 \ee
 \be\label{P*-via-vertex-tau}
\psi_n^{*(1)}(x,{\bf t})=(-1)^Ne^{-V(x,{\bf
t}^{(1)})}\frac{\langle N-1,-N|{\bar f}^{(1)}(x) g({\bf
t})|0,0\rangle}{\langle N,-N| g({\bf
t})|0,0\rangle}=x^{-N}e^{-V(x,{\bf t}^{(1)})}
\frac{D_1(x)^{-1}\tau_N({\bf t})}{\tau_{N}({\bf t})}
 \ee
 \be\label{S*-via-vertex-tau}
\psi_n^{(2)}(y,{\bf t})=e^{V(y,{\bf t}^{(2)})}\frac{\langle
N,1-N|{ f}^{(2)}(y) g({\bf t})|0,0\rangle}{\langle N,-N| g({\bf
t})|0,0\rangle}=y^{-N}e^{V(y,{\bf t}^{(2)})}
\frac{D_2(y)\tau_N({\bf t})}{\tau_N({\bf t})}
 \ee
 \be\label{S-via-vertex-tau}
\psi_n^{*(2)}(y,{\bf t})=e^{-V(y,{\bf t}^{(2)})}\frac{\langle
N,-1-N|{\bar f}^{(2)}(y) g({\bf t})|0,0\rangle}{\langle N,-N|
g({\bf t})|0,0\rangle}= y^Ne^{-V(y,{\bf
t}^{(2)})}\frac{D_2(y)^{-1}\tau_N({\bf t})}{\tau_N({\bf t})}
 \ee
 The first equality in each of the relations
(\ref{P-via-vertex-tau}),(\ref{P*-via-vertex-tau}),(\ref{S*-via-vertex-tau})
and (\ref{S-via-vertex-tau})  is an example of the evaluation of
$I_N$. Formulae (\ref{P*-via-vertex-tau}) and
(\ref{S-via-vertex-tau}) fit into the case $N_2\ge N_1\ge 0$ and
are just particular cases of formula (\ref{mainintegral-ref-1a},
see the Examples following (\ref{mainintegral-ref-1a})). The
second equalities in each of the relations
(\ref{P-via-vertex-tau})-(\ref{S-via-vertex-tau}) are examples of
bosonization formula (\ref{bosonization}).

Notice that different examples of $I_N$,
 \be\label{K11A}
K_{11}(\xi,\eta,{\bf t})=\frac{\langle N,-N|f^{(1)}(\xi){\bar
f}^{(1)}(\eta) g({\bf t})|0,0\rangle}{\langle N,-N| g({\bf
t})|0,0\rangle}=
\frac{\xi^N\eta^{-N}D_1(\xi)D_1^{-1}(\eta)\tau_N({\bf
t})}{\tau_N({\bf t})},
 \ee
 \be\label{K22A}
K_{22}(\mu,\zeta,{\bf t})=\frac{\langle N,-N|f^{(2)}(\mu){\bar
f}^{(2)}(\zeta) g({\bf t})|0,0\rangle}{\langle N,-N| g({\bf
t})|0,0\rangle}= \frac{\mu^{-N}\zeta^N
D_2(\mu)D_2^{-1}(\zeta)\tau_N({\bf t})}{\tau_N({\bf t})},
 \ee
 \be\label{K12A}
K_{12}(\xi,\zeta,{\bf t})=\frac{\langle N+1,-N-1|f^{(1)}(\xi){\bar
f}^{(2)}(\zeta) g({\bf t})|0,0\rangle}{\langle N,-N| g({\bf
t})|0,0\rangle}= \frac{\xi^N\zeta^N
D_1(\xi)D_2^{-1}(\zeta)\tau_N({\bf t})}{\tau_N({\bf t})},
 \ee
 \be\label{K21A}
K_{21}(\mu,\eta,{\bf t})=\frac{\langle N-1,-N+1|f^{(2)}(\mu){\bar
f}^{(1)}(\eta) g({\bf t})|0,0\rangle}{\langle N,-N| g({\bf
t})|0,0\rangle}=
\frac{\mu^{-N}\eta^{-N}D_2(\mu)D_1^{-1}(\eta)\tau_N({\bf
t})}{\tau_N({\bf t})},
 \ee
 (where for the sake of brevity
we omit ${\bf t}^{(1)},{\bf t}^{(2)}$-dependence in the l.h.
sides), may be considered as 2-component analogue of a modified
Cauchy-Baker-Akhiezer kernel, introduced in \cite{GO},\cite{GO-FA}
to present an explicit version of Segal-Wilson construction to
study Virasoro deformations of tau functions for the
quasi-periodical solutions of the KP (and actually for the TL)
hierarchies.

As we have obtained (see Examples following
(\ref{mainintegral-ref-1a})):

\be \ds{\mathop{ K_{11}}^{\!J}}(\mu, \zeta) = \sum_{n=0}^{J-1}
{\tilde P}_n(\mu) S_n(\zeta) + {1\over \zeta- \mu}
\label{wtK22-paper3'A}\\
 \ee
 \be
 \ds{\mathop{ K_{22}}^{\!J}}(\xi, \eta) =
\sum_{n=0}^{J-1} P_n(\xi) {\tilde S_n}(\eta) + {1\over \xi - \eta}
\label{wtK11-paper3'A}\\
 \ee
 \be \ds{\mathop{ K_{21}}^{\!J}}(\xi, \zeta) =
\sum_{n=0}^{J-1} P_n(\xi) S_n(\zeta)
\label{K12-paper3'A}\\
\ee
 \be \ds{\mathop{ K_{12}}^{\!J}}(\mu, \eta) =
\sum_{n=0}^{J-1} {\tilde P}_n(\mu)  {\tilde S}_n(\eta) - \int
{d\mu(x,y) \over (\eta-x)(\mu-y)} \label{wtK21-paper3A}
 \ee

\end{document}